\documentclass[preprint2,numberedappendix]{emulateapj-rtx4}
\usepackage{graphicx,times}
\usepackage{bm,url}
\graphicspath{{./fig/}{./png/}}

\newcommand{\bec}[1]{\mbox{\boldmath $ #1$}}
\newcommand{\EQ}{\begin{equation}}
\newcommand{\EN}{\end{equation}}
\newcommand{\EQA}{\begin{eqnarray}}
\newcommand{\ENA}{\end{eqnarray}}

\newcommand{\Eq}[1]{Equation~(\ref{#1})}

\newcommand{\Sec}[1]{Section~\ref{#1}}

\newcommand{\Fig}[1]{Figure~\ref{#1}}

\newcommand{\Figs}[2]{Figures~\ref{#1} and \ref{#2}}

\newcommand{\mean}[1]{\overline #1}

{}
{}
{}

{}
{}
\newcommand{\meanEMF}{\overline{\mbox{\boldmath ${\cal E}$}}{}}{}
{}
{}
{}
{}
{}
\newcommand{\meanBB}{\overline{\mbox{\boldmath $B$}}{}}{}
{}
{}
{}
{}
{}
{}
{}
{}

{}
{}
{}

{}

{}
{}

\newcommand{\kk}{\bm{k}}

\newcommand{\uu}{\mbox{\boldmath $u$} {}}
\newcommand{\UU}{\mbox{\boldmath $U$} {}}

\newcommand{\bb}{\mbox{\boldmath $b$} {}}
\newcommand{\BB}{\mbox{\boldmath $B$} {}}

\newcommand{\JJ}{\mbox{\boldmath $J$} {}}

\newcommand{\AAA}{\mbox{\boldmath $A$} {}}

\newcommand{\nab}{\mbox{\boldmath $\nabla$} {}}

\newcommand{\oo}{\mbox{\boldmath $\omega$} {}}

\newcommand{\SSSS}{\mbox{\boldmath ${\sf S}$} {}}

\newcommand{\sgn}{{\rm sgn}  \, {}}
\newcommand{\DD}{{\rm D} {}}

\newcommand{\dd}{{\rm d} {}}

\def\Lu{\mbox{\rm Lu}}

\def\cs{c_{\rm s}}

\def\vA{v_{\rm A}}
\def\vAz{v_{\rm A0}}

\def\kf{k_{\rm f}}

\def\Urms{U_{\rm rms}}
\def\urms{u_{\rm rms}}

\def\etat{\eta_{\it t}}
\def\etaT{\eta_{\it T}}

\def\half{{\textstyle{1\over2}}}

\def\onethird{{\textstyle{1\over3}}}

\newcommand{\s}{\,{\rm s}}

\newcommand{\cm}{\,{\rm cm}}

\newcommand{\yapj}[3]{ #1, {ApJ,} {#2}, #3}

\newcommand{\yan}[3]{ #1, {Astron.\ Nachr.,} {#2}, #3}

\newcommand{\yana}[3]{ #1, {A\&A,} {#2}, #3}

\newcommand{\yass}[3]{ #1, {Ap\&SS,} {#2}, #3}
\newcommand{\ygafd}[3]{ #1, {Geophys.\ Astrophys.\ Fluid Dyn.,} {#2}, #3}

\newcommand{\yjfm}[3]{ #1, {J.\ Fluid Mech.,} {#2}, #3}

\newcommand{\yjetp}[3]{ #1, {Sov.\ Phys.\ JETP,} {#2}, #3}
\newcommand{\yssr}[3]{ #1, {Space Sci. Rev.,} {#2}, #3}

\newcommand{\yprt}[3]{ #1, {Phys.\ Rep.,} {#2}, #3}
\newcommand{\yprl}[3]{ #1, {Phys.\ Rev.\ Lett.,} {#2}, #3}

\newcommand{\ymn}[3]{ #1, {MNRAS,} {#2}, #3}

\newcommand{\yprd}[3]{ #1, {Phys.\ Rev.\ D,} {#2}, #3}
\newcommand{\ypre}[3]{ #1, {Phys.\ Rev.\ E,} {#2}, #3}

\newcommand{\yjour}[4]{ #1, {#2}, {#3}, #4}

\newcommand{\ybook}[3]{ #1, {#2} (#3)}

\begin{document}
\title{Cosmic ray current-driven turbulence and mean-field dynamo effect}

\author{Igor Rogachevskii$^{1,2}$, Nathan Kleeorin$^{1,2}$,
Axel Brandenburg$^{2,3}$, and David Eichler$^4$}
\affil{
$^1$Department of Mechanical Engineering,
Ben-Gurion University of the Negev, POB 653, Beer-Sheva 84105, Israel\\
$^2$NORDITA, Royal Institute of Technology and Stockholm University,
Roslagstullsbacken 23, SE-10691 Stockholm, Sweden\\
$^3$Department of Astronomy, AlbaNova University Center, Stockholm University, SE 10691 Stockholm, Sweden\\
$^4$Department of Physics, Ben-Gurion University of the Negev, POB 653, Beer-Sheva 84105, Israel
}

\date{~$ $Revision: 1.246 $ $}
\begin{abstract}
We show that an $\alpha$ effect is
driven by the cosmic ray Bell instability exciting
left--right asymmetric turbulence.
Alfv\'en waves of a preferred polarization have
maximally helical motion,
because the transverse motion of each mode is
parallel to its curl.
We show how large-scale Alfv\'en modes,
when rendered unstable by cosmic ray streaming,
can create new net flux over any finite region, in the
direction of the original large-scale field.
We perform direct numerical simulations (DNS) of an
MHD fluid with a forced cosmic ray current and
use the test-field method to determine the
$\alpha$ effect and the turbulent magnetic diffusivity.
As follows from DNS, the dynamics of
the instability has the following stages: (i) in
the early stage, the small-scale Bell instability
that results in a production of small-scale
turbulence is excited; (ii) in the intermediate
stage, there is formation of larger-scale magnetic structures;
(iii) finally,  quasi-stationary large-scale turbulence is formed
at a growth rate that is comparable to that expected
from the dynamo instability,
but its amplitude over much longer timescales remains unclear.
The results of DNS are in good agreement with the theoretical estimates.

It is suggested that this dynamo is what gives
weakly magnetized relativistic shocks such as those from gamma ray
bursts a macroscopic correlation length.
It may also be important for
large-scale magnetic field amplification
associated with cosmic ray production and
diffusive shock acceleration in supernova
remnants (SNR) and blast waves from gamma ray bursts.
Magnetic field amplification by Bell
turbulence in SNR is found to be significant,
but it is limited owing to the finite time
available to the super-Alfv\'enicly expanding remnant.
The effectiveness of the mechanisms
is shown to be dependent on the shock velocity.
Limits on magnetic field growth in
longer-lived systems, such as the Galaxy
and unconfined intergalactic cosmic rays is also
discussed.
\end{abstract}


\maketitle

\section{Introduction}
\label{Introduction}

Astrophysical blast waves are strongly suspected of
amplifying the ambient magnetic field into
which they propagate.
Supernova remnants (SNR), given detailed models for their
ultra high energy gamma ray emission, indicate magnetic
fields that are
considerably stronger than the several $\mu$G fields
that are present in the interstellar medium.
The exact strength of SNR magnetic fields depends on how
small-scale bright spots are interpreted \citep{Pohl}.

Gamma ray bursts (GRB) afterglows, which are attributed to relativistic
blast waves, are generally best fit with a magnetic field strength
that is much higher than interstellar magnetic fields.
It has been suggested that the Weibel instability  is responsible for
the magnetic field production/amplification \citep{ML99},
but several difficulties with this proposal \citep{BE87} remain unsolved.
First, the fastest growing Weibel-unstable modes are of very small scale,
the ion plasma skin depth, and they should decay away resistively over
the hydrodynamical timescale of the blast wave.
Second, the electrons in the actual interstellar medium are already
magnetized as they enter the shock, and they should therefore
freeze the magnetic flux.
The Weibel instability, which creates new flux, should therefore be
suppressed at finite amplitudes, despite being unstable at infinitesimal
amplitudes.
It is not clear that these problems are resolved by simulations,
which cannot be run over hydrodynamical timescales, and which do not
always include initial electron magnetization.
In this regard, the Bell instability \citep{BELL04},
which treats the thermal plasma as a magnetized fluid
may be more relevant than kinetic approaches that ignore
the electron magnetization, i.e., that their gyro radii are small
compared with the relevant length scales of the system.

Cosmic ray (CR) protons above 200 TeV and iron nuclei above
3 PeV are difficult to account for with standard SNR parameters
\citep{Lagage-and-Cesarsky-1983}, and magnetic field
amplification would solve this problem if it occurs
on a large scale.
The smaller the scale of field amplification, the lower the maximum energy
of the CRs that can be accelerated by the shock
\citep{Eichler-and-Pohl-2011}.

Simulations of magnetic fields in the presence of cosmic rays
\citep{BELL00,BELL04} show magnetic field stretching.
This is accompanied by a jumbling of the field lines into a more
complicated geometry, and a smaller coherence length.
It has remained unclear whether this is merely turbulent
field line stretching or whether there is an additional mechanism.

It has recently been noted by \cite{BOE10}
that a circularly polarized Alfv\'en wave gives a net
electromotive force (EMF) along the direction of the
original magnetic field.
Because cosmic ray protons preferentially excite
Alfv\'en waves of a particular circular polarization,
they generate a net EMF along the original magnetic field.
The field growth is given by the curl of this EMF
which, for a plane wave, is ${\bm k} {\times} {\bm B}$,
yielding a growing Alfv\'en mode, whose polarization
vector ${\bm k} {\times} {\bm B}$ is perpendicular to the
original magnetic field.
To lowest order, this does not amplify the field but
merely bends it, and the question still remains as
to whether (i) the field lines are merely getting
stretched on this large scale, which would leave
the net flux through
any large-scale surface unchanged, or whether
(ii) there is
organized amplification, whereby the flux
of magnetic field lines
though a given large-scale surface is increased.

In this paper we note that the situation can
result in an $\alpha^2$ dynamo, in which a large-scale
magnetic field can grow along its original
direction and thus get amplified.
The effect results from a nonlinear coupling between the individual
waves that were demonstrated by \cite{BOE10} to grow.
However, there is a maximum size over which such
amplification can be effective. As the quantity
$\alpha$ has dimensions of velocity (in contrast
to the stretching timescale, which has dimensions
of time and which can be as short as the eddy
turnover time of the turbulence), this maximum
spatial scale is of the order of $\alpha \, T$,
where $T$ is the age of the blast wave. The
magnetic field on this scale can grow by at most
of the order of one $e$-fold over this time. On
the other hand, because the field is amplified
along its original direction, scales below the
maximum scale can grow exponentially, in contrast
to mere stretching, which would continually alter
the scale as the field lines lengthen.

The formal objective of the present work is to show that the parameter
$\alpha$, as defined in standard dynamo theory, is non-zero in the
presence of cosmic ray streaming instabilities, and to estimate its value.
This will imply a maximum  amount of growth on any given scale over any
given time interval.

The partial pressure $P^{\rm cr}= n^{\rm cr} \Gamma m_i c^2/3$
in cosmic rays in any logarithmic interval of energy above energy
$E_{\rm min}$ is about $P^{\rm cr} \sim \rho u_s^2/\ln(E_{\rm
max}/E_{\rm min})\lesssim \rho u_s^2/10$
\citep{Ellison-and-Eichler-1985}.
Here, $u_s$ is the streaming velocity,
$\rho=n_i m_i$ is the plasma density,
$n_i$ is the interstellar number density of
protons, $m_i$ is the mass of a plasma ion, and
$\Gamma$ is the CR's Lorentz factor.
Thus, the high energy CR number density is
\begin{eqnarray*}
n^{\rm cr} ={3P^{\rm cr} \over \Gamma m_i c^2}
\sim {3 \rho u_s^2 \over \Gamma m_i c^2
\ln(E_{\rm max}/E_{\rm min})} .
\end{eqnarray*}
The cosmic ray current ${\bm J}^{\rm cr} = n^{\rm cr}\, e \,
u_s$ due to CRs within a given logarithmic
interval of CR energy can then be factored into
dimensionless parameters as follows:
\begin{eqnarray}
{\cal J}\equiv{4\pi\over c} {J^{\rm cr}\over k
B}= {3\over 2} {u_s\over c} \, {P^{\rm cr} \over
B^2/8\pi} \, {eB\over k\Gamma m_i c^2}.
\label{KeyParameter}
\end{eqnarray}
We will show in the present study that ${\cal J}$ is
a key parameter that determines large-scale
magnetic field amplification. The last factor
$eB/(k\Gamma m_i c^2)$ in \Eq{KeyParameter}
is the deflection of the CR over its passage
through one scale length, $k^{-1}$, of the magnetic field.
The derivation of the Bell
instability that is caused by the CR current in
plasma, assumes that the deflection of the CR is small.
For the firehose instability, the subject of
a separate investigation, it can be larger than
unity, as the CRs would be bound to the field
lines. However, a situation can occur in which
the CRs maintain a steady anisotropy, $A$, even
if they scatter, and the choice for the streaming
velocity is then $u_s = A \, c$, and the shock velocity
is assumed to be approximately the same.

The factor $u_s/c$ is typically of the order of
$10^{-2}$ or less, both for the Galaxy as a
whole, where the anisotropy is limited by direct
measurements, and for the cosmic rays accelerated
by forward shocks of blast waves from supernovae
remnants (SNR), where the CR precursor moves
essentially at the shock velocity, typically
$10^{-3}$ to $10^{-2}$ times $c$. Very young
SNR can have somewhat higher shock velocities,
but the CR accelerated in them may then be more
prone to adiabatic losses. Note that because we
assume the streaming velocity to be the velocity
of the shock, we may allow for the possibility
that CRs scatter in our estimate of the CR current.
But we still have to limit the time
over which a given parcel of fluid is exposed to
the CR current to be less than the shock crossing
time -- the time it takes the shock to cross the
length of the CR precursor at that energy.

The factor $3P^{\rm cr} / (B^2/4\pi)$ is of the
order of unity for the Galaxy as a whole if all
relativistic CRs are included. In this case, the
anisotropy must be taken to be at most $10^{-3}$
and $u_s\lesssim 10^{-3}c$. In the case of
supernova blast waves, the factor $3P^{\rm cr} /
(B^2/4\pi)$ can be as high as $\sim u_s^2/\vA^2$,
where $\vA$ is the Alfv\'{e}n speed.
Thus, the CR pressure can be a significant fraction
of the ram pressure $\rho u_s^2$; see, e.g.,
\cite{Ellison-and-Eichler-1985}.

Altogether, we can choose a plausible value for
${\cal J}$ of $\sim 3 u_s^3/c \vA^2
\ln(E_{\rm max}/E_{\rm min})$ which,
for supernova remnants (SNR), can range from
$10^4$ for young SNR ($u_s \sim c/10\sim 10^3
\vA$) to order unity for old ones. For the Galaxy
as a whole, we are probably limited to $3 P^{\rm
cr}/ (B^2/4\pi) \lesssim 1  $.
For collisionless shock waves generated in the interstellar medium by
GRBs, the value of ${\cal J}$ can be enormous -- of
order $10^{14}$, and the $\alpha$ effect may be particularly effective
in that context.
This will be discussed in greater detail below, where
relativistic effects are included more carefully.

The physical time over which the mechanism can
operate on a given patch
of upstream fluid in the case of a
SNR is one expansion time,
or $R_s/u_s$, where $R_s$ is the radius of the
forward shock.
As $u_s$ does not appear explicitly in the
simulation parameters,
we use the above expression ${\cal J}\sim u_s^3/c \vA^2$
to substitute for $u_s$ in terms of $J^{\rm cr}$
and $\vA$.
For the Galaxy as a whole, we are limited by
the Hubble time to be about $10^4 R_{\rm G}/\vA$,
where $R_{\rm G}$ is the size of the Galaxy.

\section{Governing equations}

We consider magnetohydrodynamic (MHD) flows consisting
of background plasma ions
of number density $n_i$, electrons of number density $n_e$, and
cosmic ray protons of number density $n^{\rm cr}$.
The equation of motion in MHD flows with cosmic rays imbedded in a
background plasma, reads \citep{C10}:
\begin{eqnarray}
\rho \, {\DD {\bm U} \over \DD t} = - \bec{\nabla} P +
{1 \over c} {\bm J} {\bm \times} {\bm B}
+ e \, (n_i - n_e) \, {\bm E}
+ {\bm F}_\nu,
\label{A1}
\end{eqnarray}
where $\DD/\DD t=\partial / \partial t + {\bm U} {\bm \cdot} \bec{\nabla}$
is the advective derivative, ${\bm U}$ is the fluid velocity,
$\rho$ and $P$ are the fluid density and pressure, respectively,
${\bm E}$ and ${\bm B}$ are the electric and magnetic
fields, and ${\bm F}_\nu$ is the viscous force.
The densities of the plasma current,
${\bm J}$, and of cosmic ray protons, ${\bm J}^{\rm cr}$,
are the sources of magnetic field in Maxwell's equations:
\begin{eqnarray}
&& \bec{\nabla} {\bm \times} {\bm B} ={4 \pi \over c}
\left({\bm J} + {\bm J}^{\rm cr}\right) ,
\label{A2}\\
&&\bec{\nabla} {\bm \times} {\bm E} =
- {1 \over c} {\partial {\bm B} \over \partial t},
\label{A3}
\end{eqnarray}
with Ohm's law: $ {\bm J} = \sigma \left({\bm E}
+ c^{-1} {\bm U} {\bm \times} {\bm B} \right)$,
where $\sigma$ is the electrical conductivity of the gas,
and we have neglected in \Eq{A2} the displacement
current, because the conductivity is high and the
fluid motions slow compared with the speed of light.
These equations yield the induction equation,
\begin{eqnarray}
{\partial {\bm B} \over \partial t} = \bec{\nabla}
{\bm \times}\left({\bm U} {\bm \times} {\bm B}
- \eta \bec{\nabla} {\bm \times} {\bm B}
+ {c \, {\bm J}^{\rm cr} / \sigma} \right),
\label{A4}
\end{eqnarray}
where $\eta=c^2 /4 \pi \sigma$ is the magnetic diffusivity.
We assume quasi-neutrality for the whole system,
i.e., $n_i + n^{\rm cr} = n_e$, and \Eq{A1} reads:
\begin{eqnarray}
\rho \, {\DD{\bm U} \over\DD t} &=& - \bec{\nabla} P
+ {1 \over 4 \pi} \left(\bec{\nabla} {\bm \times}
{\bm B}\right) {\bm \times} {\bm B} + {\bm F}_\nu
- {1 \over c} {\bm J}^{\rm cr} {\bm \times} {\bm B}
\nonumber \\
&& + {1 \over c} e n^{\rm cr}
\, \left({\bm U} {\bm \times}{\bm B}\right),
\label{A5}
\end{eqnarray}
where we have used \Eq{A2} and assumed that
$|{\bm J} / \sigma| \ll
c^{-1} |{\bm U} {\bm \times} {\bm B}|$,
i.e., we consider plasma flows with large hydrodynamic and
magnetic Reynolds numbers.
Hereafter we assume that the cosmic ray velocity is much
larger than fluid velocity ${\bm U}$, so that the last term
in \Eq{A5} vanishes.
The plasma density is determined by the continuity equation:
\begin{eqnarray}
{\partial \rho \over \partial t} + \bec{\nabla}
{\bm \cdot}  \rho \, {\bm U} =0 .
\label{A6}
\end{eqnarray}

Following \cite{BELL04}, let us consider the
equilibrium: ${\bm J}^{\rm eq} + {\bm J}^{\rm
cr}=0$, ${\bm B}^{\rm eq}={\bm B}_\ast=$const,
${\bm U}^{\rm eq}=0$ and $\rho^{\rm
eq}=\rho_\ast=$const. Linearized
Eqs.~(\ref{A2})--(\ref{A6}) for small
perturbations yield the following dispersion relation:
\begin{eqnarray}
&& \gamma^2\left(\gamma \gamma_{\it B}
+ \omega_s^2\right) \left(\gamma^2+\omega_A^2\right)=
\omega_A^2 \left(\gamma \gamma_{\it B}\, {k^2 \over k_z^2}
+ \omega_s^2 \right)
\nonumber\\
&& \times \left[\Big({\bec{\omega}^{\rm cr} k_z \over k}\Big)^2 - \omega_A^2
-\gamma^2\right],
\label{AA7}
\end{eqnarray}
where $\omega_A = {\bm k} {\bm \cdot} \vA$
is the frequency of the Alfv\'{e}n waves, $\omega_s = k c_s$
is the frequency of the sound waves, $c_s$ is the sound speed,
${\bm k}$ is the wavenumber,
$\vA = {\bm B}_\ast/
\left(4 \pi \rho\right)^{1/2}$ is the Alfv\'{e}n speed,
$\omega^{\rm cr} = c^{-1} \, J^{\rm cr}\, (4 \pi /\rho)^{1/2}$,
$\gamma =\gamma_{\it B} + \eta k^2$,
$\gamma_{\it B}$ is the growth rate of an instability
and we have considered the case where the equilibrium
magnetic field ${\bm B}_\ast$
and ${\bm J}^{\rm cr}$ are directed along the $z$ axis.

In this system, the non-resonant Bell instability
\citep{BELL04,BELL00} is excited
by the cosmic ray current that causes growing MHD modes.
The growth rate of this instability for $\beta \gg 1$
that follows from \Eq{AA7}, is given by
\begin{eqnarray}
\gamma_{\it B} =   \left[{|\omega_A \,
\omega^{\rm cr} \, k_z| \over k} -\omega_A^2
\,\right]^{1/2} - \eta k^2,
\label{A7}
\end{eqnarray}
where $\beta$ is the ratio of gas pressure to magnetic
pressure.
For incompressible MHD modes $\tilde{\bm b}({\bm k})
= i ({\bm k} {\bm \cdot} {\bm B}_\ast) \,
\tilde{\bm u}({\bm k}) / \gamma_{\it B}$
\citep{BELL04}, where $\tilde{\bm u}$ and $\tilde{\bm b}$
are perturbations of velocity and magnetic field.

When $\beta \ll 1$, the growth rate of this
instability that follows from \Eq{AA7}, is given by
\begin{eqnarray}
\gamma_{\it B} &=&  {|\omega_A| \over \sqrt{2}}
\, \biggl\{\left[\left(1 - {k \over k_z}\right)^2
 + \left({2 \omega^{\rm cr}
\over \omega_A}\right)^2 \right]^{1/2}
\nonumber\\
&&  - \left(1 + {k^2 \over k_z^2}\right) \biggr\}^{1/2}
- \eta k^2.
\label{A8}
\end{eqnarray}

The nonlinear stage of this instability has been
investigated in a number of publications
\citep[see,
e.g.,][]{BELL04,BELL05,PLM06,RODK08,ZPV08,AB09,LM09,VBE09,ZE10,BOE10,BER11}.
These studies have demonstrated that this
instability produces small-scale turbulence with
cascading of turbulence energy into larger and
smaller scales.

In this paper we discuss the possibility of mean-field dynamo action caused
by the interaction of mean electric current of cosmic
ray particles with small-scale background homogeneous
turbulence produced by the Bell instability.
This paper can be considered as an extension of the recent
study by \cite{BOE10} who demonstrated,
using a multi-scale quasi-linear mean-field
approach, that small-scale Bell-type turbulence can result in the growth of
long-wavelength obliquely propagating modes \citep{BOE10}.

\section{Large-scale instability}

In this section we discuss mean-field dynamo action
in small-scale turbulence produced by the Bell
instability in a plasma with a given mean electric
current of cosmic ray ions.
The importance of this effect is determined by the ratio of
the cosmic ray current to the ambient field; see \Eq{KeyParameter}.

\subsection{Mean field dynamo equations}

We use a mean field approach in which magnetic
and velocity fields are divided into mean and
fluctuating parts, ${\bm U} = \overline{\bm U}
+ {\bm u}$ and ${\bm B} = \overline{\bm B} +
{\bm b}$, where ${\bm u}$ and ${\bm b}$ are fluctuations
of velocity and magnetic field, $\overline{\bm B}$
and $\overline{\bm U}$ are the mean magnetic and
velocity fields, and the fluctuating fields
have zero mean values.
We consider the case when magnetic and fluid Reynolds
numbers are large.
This implies that the nonlinear terms in the induction
and Navier-Stokes equations are much larger than
the dissipative and viscous terms.
In this case the quasi-linear approach for
determining the turbulent transport coefficients
(e.g., the $\alpha$ effect and turbulent magnetic diffusivity)
does not work. We use instead the spectral $\tau$ relaxation
approximation \citep{O70,PFL76,KRR90,RK04}
that is valid for large magnetic and fluid Reynolds numbers.
A justification for the $\tau$ approximation in
different situations has been performed in numerical
simulations and analytical studies
\cite[see, e.g.,][]{BS05,RKB11}. For more details see
Appendix \ref{Tau}.

Averaging Eqs.~(\ref{A4}) and~(\ref{A5}) over an ensemble
of turbulent eddies yields the following mean-field equations:
\begin{eqnarray}
{\partial \overline{\bm B} \over \partial t} &=&
\bec{\nabla} {\rm \times} \left(\overline{\bm U}
{\bm \times} \overline{\bm B} + \overline{{\bm u}
{\bm \times} {\bm b}} - \eta \bec{\nabla}
{\bm \times} \overline{\bm B} \right),
\label{D1}\\
\overline{\rho} \, {\DD\overline{\bm U} \over\DD t}
&=& - \bec{\nabla} \overline{P}
+ {1 \over 4 \pi} \left(\bec{\nabla} {\bm \times} \overline{\bm B}
\right) {\bm \times} \overline{\bm B}
- {1 \over c} \overline{{\bm J}^{\rm cr}} {\bm \times}
\overline{\bm B}
\nonumber \\
&& + {1 \over c} e n^{\rm cr} \, \left(\overline{\bm U}
{\bm \times} \overline{\bm B}\right) - \nabla_j \overline{
{\bm u} u_j} + \overline{{\bm F}_\nu},
\label{D2}
\end{eqnarray}
where $\bec{\cal E}(\overline{\bm B}) = \overline{{\bm u}
\times {\bm b}}$ is the mean electromotive force,
$\overline{{\bm J}^{\rm cr}}$ is the mean density of
the electric current of cosmic ray particles.
For large hydrodynamic and magnetic Reynolds numbers
we can neglect kinematic viscosity, $\nu$, and magnetic diffusivity,
$\eta$, in comparison with the turbulent viscosity and
turbulent magnetic diffusion.

\subsection{Contributions to the $\alpha$ effect}

We will show in this study that, formally, there
are two contributions to the $\alpha$ effect
caused by: (i) existing kinetic helicity produced
by the Bell-instability (referred to
as $\overline{{\bm u}^{(0)} \cdot (\bec{\nabla}
\times {\bm u}^{(0)})}$ below; (ii) correlations
in the forcing by the mean cosmic ray current in
the presence of small-scale anisotropic
magnetic fluctuations which create
further perturbation  ${\bm u}^{(1)}$ in the velocity
field ${\bm u}^{(0)}$. Effect (ii) causes
opposite sides of a magnetic loop to be  forced
in opposite directions, thereby twisting the loop
out of its original plane.  The distinction
between the two terms may be more formal than
physical.
That is, in an unstable circularly polarized
Alfv\'en mode, the additional helicity added by the
cosmic ray forcing term is merely a continuation
of the process that formed the already existing
helicity.

The $\alpha$ effect, however, is distinct from
the linear growth of the circularly polarized
Alfv\'en wave. It can be thought of as one
circularly polarized Alfv\'en wave riding on
another, somewhat longer wavelength Alfv\'en wave
(which need not be circularly polarized). The
longer wave makes a perpendicular component to
the original field, while the stretching of the
perpendicular component  (into a loop, say)
together with  the twisting of the loop  restores
newly created flux back into the original
direction. It is the {\it nonlinear coupling} of
two waves, each of which is of the sort
discussed by \cite{BOE10}. This coupling
can scatter energy in the two modes of
wavenumbers $k_1$, $k_2$ into modes of much
longer wavelength, and thereby amplify the field
at large scale in its original direction.

Let us consider the case where the equilibrium
uniform mean magnetic field $\overline{\bm
B}_\ast$ and the mean density of the electric
current of accelerated particles $\overline{{\bm
J}^{\rm cr}}$ are directed along the $z$ axis. We
take into account effects which are linear in
perturbations of the mean magnetic field:
$\tilde{\bm B}=\overline{\bm B} -\overline{\bm
B}_\ast$, i.e., we consider kinematic mean-field dynamo.

The first contribution to the $\alpha$ effect is
caused by the helical part of the turbulence.
A non-zero kinetic helicity is caused by the
Bell instability that results in the production of
small-scale helical turbulence.
This contribution to the mean electromotive force
is given by ${\cal E}_{i}^{\rm(I)} = \alpha_{ij}^{\rm(I)}
\, \tilde{B}_j+...$
(see Appendix \ref{Alpha}),
where $\alpha_{ij}^{\rm(I)}
= \alpha^{\rm cr}_1 \, \delta_{ij}$ is an isotropic $\alpha$ effect,
\begin{eqnarray}
\alpha^{\rm cr}_1  &=& - C_1 \, \tau_0 \,
\overline{{\bm u}^{(0)} \cdot
(\bec{\nabla} \times {\bm u}^{(0)})} ,
\label{D3}
\end{eqnarray}
with coefficient $C_1 \approx 1/3$, and dots
referring to higher-order terms that will be
considered later in \Sec{Test_field_method}.
In the rest of this section the dots will not be noted explicitly.
Here, ${\bm u}^{(0)}$ are the velocity fluctuations of
the background turbulence.
Equation~(\ref{D3})
is a well-known result for the $\alpha$ effect caused by kinetic
helicity of the turbulence
\citep[see][]{KR80,M78,P79,ZRS83}.

The second contribution to the $\alpha$ effect
is caused by incompressible and non-helical parts
of the anisotropic turbulence
interacting with the mean cosmic ray current.
In particular, this contribution to the mean
electromotive force
is given by ${\cal E}_{i}^{\rm(II)} = \alpha_{ij}^{\rm(II)}
\, \tilde{B}_j$, where $\alpha_{ij}^{\rm(II)}=
\alpha^{\rm cr}_2 \, (\delta_{ij} + e_{i} e_{j})$
(see Appendix \ref{Alpha}),
\begin{eqnarray}
\alpha^{\rm cr}_2 &=& C_2 \, \left({4\pi\over c}
{\overline{J^{\rm cr}} \, \ell_0 \over \overline{B}_\ast} \right)^{1/2}
\, \overline{V}_A
\, \, \sgn{\left(\overline{{\bm J}^{\rm cr}} \cdot {\bm B}_\ast\right)},
\label{D4}
\end{eqnarray}
$\overline{V}_A=\overline{B}_\ast / (4 \pi \overline{\rho})^{1/2}$
is the mean Alfv\'{e}n speed
based on the equilibrium mean magnetic
field $\overline{\bm B}_\ast$,
${\bm e}$ is the unit vector directed along
$\overline{\bm B}_\ast$,
$C_2 = 4 \, \epsilon \,  (q-1) / 3(2q-3) $, $0 <
\epsilon \leq 1$ is the anisotropy parameter of
small-scale turbulence and $q$ is the exponent of
the energy spectrum of the turbulence.
Note that $C_2 \approx 8/3$ for $\epsilon \approx 1$ and $q=5/3$.
The total $\alpha$ effect is the sum of the two contributions:
\begin{eqnarray}
\alpha_{ij}^{\rm cr} = \alpha_{ij}^{\rm(I)} + \alpha_{ij}^{\rm(II)}.
\label{D27}
\end{eqnarray}

The mechanism of the second contribution,
$\alpha_{ij}^{\rm(II)}$, to the $\alpha$ effect can
be understood by the following reasoning.
Tangling of the mean magnetic field $\tilde{\bm
B}=\overline{\bm B} -\overline{\bm B}_\ast$ by
velocity fluctuations of the background
anisotropic turbulence ${\bm u}^{(0)}$ produces
magnetic fluctuations:
\begin{eqnarray}
{\partial {\bm b}^{(1)} \over \partial t} \propto
(\tilde{\bm B} {\bf \cdot} \bec{\nabla}) {\bm u}^{(0)}.
\label{EX1}
\end{eqnarray}
The generated magnetic fluctuations ${\bm b}^{(1)}$
interacting with the cosmic ray current
$\overline{{\bm J}^{\rm cr}}$ produce additional
velocity fluctuations:
\begin{eqnarray}
{\partial {\bm u}^{(1)} \over \partial t} \propto
- {1 \over c \, \overline{\rho}} \overline{{\bm J}^{\rm cr}}
{\bm \times}{\bm b}^{(1)}.
\label{EX2}
\end{eqnarray}
These velocity fluctuations contribute to the mean
electromotive force $\bec{\cal E}^{\rm(II)}
= \overline{{\bm u}^{(1)}\times {\bm b}^{(0)}}$.
Here, ${\bm b}^{(0)}$ are the magnetic fluctuations
resulting directly from the Bell instability,
just like the velocity perturbations ${\bm u}^{(0)}$.
In particular, the stretching of the original magnetic
field determined by \Eq{EX1}, and rotation of
the stretched magnetic loop determined by \Eq{EX2}
create an electric field $\bec{\cal E}^{\rm(II)}$ along the
original magnetic field $\tilde{\bm B}$.
The Lorentz force in \Eq{EX2} plays the role of rotation
of a magnetic loop.
Using Eqs.~(\ref{EX1}) and (\ref{EX2}), we can estimate
the mean electromotive force $\bec{\cal E}^{\rm(II)}$
using dimensional reasoning.
Indeed, the velocity fluctuations can be estimated as
\begin{eqnarray}
{\bm u}^{(1)} \propto
- {\tau \over c \, \overline{\rho}}
\overline{{\bm J}^{\rm cr}}
{\bm \times}{\bm b}^{(1)}
\propto
- {\tau^2 \over c \, \overline{\rho}} \,
\overline{{\bm J}^{\rm cr}}{\bm \times}(\tilde{\bm B}
{\bf \cdot} \bec{\nabla}) {\bm u}^{(0)},
\label{EX3}
\end{eqnarray}
where $\tau$ is the characteristic time of the turbulence.
Therefore, the mean electromotive force
$\bec{\cal E}^{\rm(II)}$ is estimated as
\begin{eqnarray}
{\cal E}^{\rm(II)}_i \!\propto\! {\tau^2 \over c \,
\overline{\rho}} \left(\overline{J_i^{\rm cr}} \, \, \,
\overline{b_n^{(0)} \nabla_j u_n^{(0)}} -
\overline{J_n^{\rm cr}} \, \, \, \overline{b_n^{(0)}
\nabla_j u_i^{(0)}} \right) \tilde B_j.\;
\label{EX4}
\end{eqnarray}
The mean electromotive force $\bec{\cal E}^{\rm(II)}$
is proportional to the mean magnetic field $\tilde{\bm B}$,
i.e., ${\cal E}^{\rm(II)}_i=a_{ij} \, \tilde B_j$ and
the tensor $\alpha_{ij}^{\rm(II)}$ is the symmetric part
of $a_{ij}$.
Therefore,
\begin{eqnarray}
\alpha_{ij}^{\rm(II)} &\propto& {\tau^2 \over 2 c \,
\overline{\rho}} \biggl(\overline{J_i^{\rm cr}} \, \, \,
 \overline{b_n^{(0)} \nabla_j u_n^{(0)}} + \overline{J_j^{\rm cr}}
\, \, \, \overline{b_n^{(0)} \nabla_i u_n^{(0)}}
\nonumber\\
&& - \overline{J_n^{\rm cr}} \, \, \, \overline{b_n^{(0)}
\nabla_j u_i^{(0)}}
- \overline{J_n^{\rm cr}} \, \, \, \overline{b_n^{(0)}
\nabla_i u_j^{(0)}} \biggr) .
\label{EX5}
\end{eqnarray}
It follows from \Eq{EX5} that for
isotropic background turbulence,
$\overline{b_i^{(0)} u_j^{(0)}} \propto
\delta_{ij}$, and $\alpha_{ij}^{\rm(II)}$ vanishes.
Note also that in many kinds of background
turbulence,  the tensor $\overline{b_i^{(0)}
u_j^{(0)}}$ as well as the background mean
electromotive force $\bec{\cal E}^{(0)}= {\bm
u}^{(0)}{\bm \times}{\bm b}^{(0)}$ vanish. On the
other hand, for the turbulence produced by the
Bell instability, the tensor $\overline{b_i^{(0)}
u_j^{(0)}}$ does not vanish.

Now let us compare this mechanism for the $\alpha$ effect
with the standard mechanism based on nonzero kinetic helicity.
The mean electromotive force $\bec{\cal E}^{\rm(I)}
= \overline{{\bm u}^{(0)}\times {\bm b}^{(1)}}$, where
the magnetic fluctuations ${\bm b}^{(1)}$ are generated by the
tangling of the mean magnetic field $\tilde{\bm B}$
by the velocity fluctuations of the background
turbulence ${\bm u}^{(0)}$.
Therefore, the mean electromotive force is given by
\begin{eqnarray}
\bec{\cal E}^{\rm(I)} \propto \tau \overline{{\bm u}^{(0)}
\times (\tilde{\bm B} {\bf \cdot} \bec{\nabla}) {\bm u}^{(0)}} ,
\label{EX6}
\end{eqnarray}
and the $\alpha_{ij}^{\rm(I)}$ tensor is
\begin{eqnarray}
\alpha_{ij}^{\rm(I)} &\propto& \tau \biggl(\varepsilon_{inm}
\, \overline{u_m^{(0)} \nabla_j u_n^{(0)}} + \varepsilon_{jnm}
\, \overline{u_m^{(0)} \nabla_i u_n^{(0)}} \biggr) ,
\label{EX7}
\end{eqnarray}
where the kinetic helicity is produced by the Bell instability
and $\chi_{ij}=\varepsilon_{inm}
\, \overline{u_m^{(0)} \nabla_j u_n^{(0)}} + \varepsilon_{jnm}
\, \overline{u_m^{(0)} \nabla_i u_n^{(0)}}$ is the symmetric
tensor of the kinetic helicity for anisotropic turbulence.
Using the model for helical background turbulence \Eq{EX7}
can be reduced to \Eq{D3}.
Therefore the $\alpha_{ij}^{\rm(I)}$ tensor is directly related to the kinetic helicity.
Now let us estimate the kinetic helicity for the Bell background turbulence.
The contribution of the cosmic ray current to the velocity of the background turbulence is
\begin{eqnarray}
{\partial {\bm u}^{(0)} \over \partial t} \propto
- {1 \over c \, \overline{\rho}} \overline{{\bm J}^{\rm cr}}
{\bm \times}{\bm b}^{(0)}.
\label{EX8}
\end{eqnarray}
Therefore, the corresponding contribution of the cosmic ray
current to the vorticity of the background turbulence is
\begin{eqnarray}
{\partial  \over \partial t} \bec{\nabla} \times {\bm u}^{(0)}
\propto{1 \over c \, \overline{\rho}} \,
\left(\overline{{\bm J}^{\rm cr}} {\bf \cdot} \bec{\nabla}
\right) {\bm b}^{(0)}.
\label{EX9}
\end{eqnarray}
Multiplying \Eq{EX8} by the vorticity
$\bec{\nabla} \times {\bm u}^{(0)}$ and
\Eq{EX9} by the velocity ${\bm u}^{(0)}$,
and averaging over the ensemble we obtain:
\begin{eqnarray}
{\partial  \over \partial t} \overline{{\bm u}^{(0)} {\bf \cdot}
\left(\bec{\nabla} \times {\bm u}^{(0)}\right)} \propto
&& -
{\overline{J_j^{\rm cr}} \over c \, \overline{\rho}} \, \Big(
\overline{{\bm b}^{(0)} \times \left(\bec{\nabla} \times {\bm u}^{(0)}\right)_j}
\nonumber\\
&&- \overline{u_n^{(0)} \nabla_j b_n^{(0)}}
\Big).
\label{EX10}
\end{eqnarray}
Dimensional reasoning yields the following estimate for the
the kinetic helicity for the Bell background turbulence:
\begin{eqnarray}
\overline{{\bm u}^{(0)} {\bf \cdot} \left(\bec{\nabla}
\times {\bm u}^{(0)}\right)} \propto
&& \left({\tau^2 \, \overline{J_j^{\rm cr}} \,
\overline{J_i^{\rm cr}} \over c^2 \, \overline{\rho}^2}
\right) \, \, \varepsilon_{jnm}
\, \overline{b_m^{(0)} \nabla_i b_n^{(0)}}
\nonumber\\
&&+ {\tau \, \overline{J_j^{\rm cr}} \over c \,
\overline{\rho}} \, \overline{u_n^{(0)} \nabla_j b_n^{(0)}},
\label{EX11}
\end{eqnarray}
where we have taken into account that $\bec{\nabla} \times
{\bm u}^{(0)} \propto (\tau / c \, \overline{\rho})
\, (\overline{{\bm J}^{\rm cr}} {\bf \cdot} \bec{\nabla}) {\bm b}^{(0)}$.
The first term in the right hand side of \Eq{EX11}
is related to the current and magnetic helicities.
For the Bell instability ${\bm b}^{(0)}({\bm k})
\approx i ({\bm k} {\bm \cdot} {\bm B}_\ast) \,
{\bm u}^{(0)}({\bm k}) / \gamma_{\it B}$.
Therefore, Eqs.~(\ref{EX5}) and~(\ref{EX11}) show that only
part of the contribution to
$\alpha_{ij}^{\rm(II)}$ may be related to the kinetic helicity.
This contribution is caused by the last term in the right hand
side of \Eq{EX11}.
Note that \Eq{EX11} for the kinetic helicity
is different from Equation~(C6) of \cite{BOE10}, where no substitution
for ${\bm b}^{(1)}$, as in \Eq{EX3}, is performed.

\subsection{Large-scale instability for $\tilde{\bm B}(t, x, y)$}

Let us start to analyze the large-scale instability using
a case of incompressible flow when perturbations
of velocity and magnetic field are independent of $z$.
In this case
perturbations of the mean magnetic field $\tilde{\bm B}(t, x, y)$ are
determined by the following equation:
\begin{eqnarray}
{\partial \tilde{\bm B} \over \partial t} =
(\overline{\bm B}_\ast {\rm \cdot} \bec{\nabla})
\tilde{\bm U} + \bec{\nabla} {\rm \times} \left(\overline{
{\bm u} {\bm \times} {\bm b}}\right) \,,
\label{D7}
\end{eqnarray}
where $\left(\overline{ {\bm u} {\bm \times} {\bm b} }\right)_i
= \alpha^{\rm cr}_{ij} \, \tilde{B}_j - \etat
\, (\bec{\nabla} {\rm \times} \tilde{\bm B})_i$,
$\, \etat = C_\eta \, u_0 \, \ell_0$ is the
turbulent magnetic diffusivity,
$C_\eta \approx 1/3 $ is a constant, $\ell_{0}$ is
the maximum (integral) scale of turbulent motions,
$u_{0}$ is the characteristic turbulent velocity in
the integral scale of turbulence,  $\overline{\bm B}
= \overline{\bm B}_\ast + \tilde{\bm B}$ and
$\tilde{\bm U}(t, x, y)$ are the perturbations of
the mean velocity field.
Here, for simplicity, we neglect small anisotropic
contributions to the turbulent magnetic diffusion.
Since the equilibrium uniform mean magnetic field
$\overline{\bm B}_\ast$ is directed along $z$ axis,
the first term, $(\tilde{\bm B}_\ast {\rm \cdot}
\bec{\nabla}) \tilde{\bm U}$, on the right hand
side of \Eq{D7} vanishes for perturbations
which are independent of $z$.
In this case the mean-field induction equation~(\ref{D7})
is decoupled from the mean-field momentum equation.
The perturbations of the mean magnetic field $\tilde{\bm B}$
can be written in the form of the axisymmetric field:
$\tilde{\bm B}= \tilde{B}_z(t,x,y) \, {\bm e}
+ \bec{\nabla} {\rm \times} [\tilde{A}(t,x,y) \, {\bm e}]$.

Let us start the analysis with the simpler case in which
the tensor $\alpha^{\rm cr}_{ij} = \alpha^{\rm cr}
\,\delta_{ij}$ is isotropic.
Then the functions $\tilde{B}_z(t,x,y)$ and
$\tilde{A}(t,x,y)$ are determined by the following
equations which follow from \Eq{D7}:
\begin{eqnarray}
{\partial \tilde{B}_z \over \partial t} &=&
- \alpha^{\rm cr} \Delta \tilde{A}
+ \etat \, \Delta_{\rm H} \tilde{B}_z ,
\label{D8}\\
{\partial \tilde{A} \over \partial t} &=&
\alpha^{\rm cr} \tilde{B}_z + \etat
\, \Delta_{\rm H} \tilde{A} ,
\label{D9}
\end{eqnarray}
where $\Delta_{\rm H} = \nabla_x^2 + \nabla_y^2$.
We seek a solution of the mean-field dynamo equations (\ref{D8})
and (\ref{D9}) of the form
$ \propto \exp[\gamma_{\rm inst} \, t + i (K_x \, x + K_y \, y)]$.
The growth rate $\gamma_{\rm inst}$ of the
mean-field dynamo instability in homogeneous turbulent
plasma with cosmic rays is then given by
\begin{eqnarray}
\gamma_{\rm inst} = |\alpha^{\rm cr}  \, K|
- \etat \, K^2
\;,
\label{D10}
\end{eqnarray}
where $K^2 = K_x^2 + K_y^2$ and $\alpha^{\rm cr}
=\alpha^{\rm cr}_1$.
This large-scale dynamo instability is called $\alpha^2$
dynamo, because this dynamo is caused by the interaction
between two modes, toroidal, $\tilde{\bm B}^{(T)}
= \tilde{B}_z(t,x,y) \, {\bm e}$, and poloidal
mean magnetic fields, where poloidal field,
$\tilde{\bm B}^{(P)} = \bec{\nabla} {\rm \times}
[\tilde{A}(t,x,y) \, {\bm e}]$ is determined only
by the potential $\tilde{A}(t,x,y)$.
The toroidal field is generated from the poloidal
field by the $\alpha$ effect due to the first term,
$- \alpha^{\rm cr} \Delta \tilde{A}$ in \Eq{D8},
while the poloidal field is generated from the toroidal
field by the $\alpha$ effect [due to the first term,
$\alpha^{\rm cr} \Delta \tilde{B}_z$ in \Eq{D9}].
This implies that the $\alpha$ effect acts twice
in the positive feedback loop, and these
interactions between the magnetic field components cause
the $\alpha^2$ mean-field dynamo.

Now we consider the case in which the tensor
$\alpha^{\rm cr}_{ij}$ is anisotropic.
In this case the functions $\tilde{B}_z(t,x,y)$ and
$\tilde{A}(t,x,y)$ are given by:
\begin{eqnarray}
{\partial \tilde{B}_z \over \partial t} &=&
- \alpha^{\rm cr}_{yy} \nabla^2_x \tilde{A}
- \alpha^{\rm cr}_{xx} \nabla^2_y \tilde{A}
+ \etat \, \Delta_{\rm H} \tilde{B}_z ,
\label{D15}\\
{\partial \tilde{A} \over \partial t} &=&
\alpha^{\rm cr}_{zz} \tilde{B}_z + \etat
\, \Delta_{\rm H} \tilde{A} ,
\label{D16}
\end{eqnarray}
where $\alpha^{\rm cr}_{xx}=\alpha^{\rm cr}_{yy}
\not=\alpha^{\rm cr}_{zz}$.
The growth rate $\gamma_{\rm inst}$ of the
mean-field dynamo instability in this case is
\begin{eqnarray}
\gamma_{\rm inst} = |K| \, \sqrt{\alpha^{\rm cr}_{yy} \,
\alpha^{\rm cr}_{zz}} - \etat \, K^2 ,
\label{D17}
\end{eqnarray}
and the ratio of magnetic energies along and perpendicular
to the direction of the cosmic ray current is given by
\begin{eqnarray}
{\tilde{B}_z^2 \over \tilde{B}_x^2+\tilde{B}_y^2}=
{\alpha^{\rm cr}_{xx} \over \alpha^{\rm cr}_{zz}} ,
\label{D18}
\end{eqnarray}
where $\alpha^{\rm cr}_{xx}=\alpha^{\rm cr}_{yy}
=\alpha^{\rm cr}_1 +\alpha^{\rm cr}_2$ and
$\alpha^{\rm cr}_{zz}=\alpha^{\rm cr}_1
+2\alpha^{\rm cr}_2$.

Comparing with \cite{BOE10}, who studied a long-wave instability for
incompressible flows of a plasma with cosmic rays, we note that their
growth rate vanishes when perturbations of mean magnetic and velocity
fields are independent of $z$; see their Equation~(23).
In this sense, the mean-field dynamo mechanism
studied in our paper for $K_z=0$ is a complementary
effect to a mechanism related to the long-wave
instability discussed by \cite{BOE10}.

\subsection{Large-scale instability for $\tilde{\bm B}(t, z)$}

Now let us consider the case when perturbations of
mean magnetic and velocity fields depend only on $z$.
The perturbations $\tilde{B}_z=0$ (because
${\bm \nabla} \cdot \tilde{\bm B}=0$), and
the large-scale dynamo instability can generate
magnetic field only in the direction perpendicular
to the cosmic ray current.
In this case we seek  a solution of the linearized mean-field dynamo
equations~(\ref{D1}) and~(\ref{D2})
of the form $ \propto \exp(\gamma_{\rm inst}
\, t + i K_z \, z)$.
The growth rate $\gamma_{\rm inst}$ of the large-scale
dynamo instability is given by
\begin{eqnarray}
\gamma_{\rm inst} &=& \left[\left|\Omega_A \,
\Omega^{\rm cr}\right| -\Omega_A^2 + {1 \over 4}
\left(\alpha^{\rm cr} K_z\right)^2 \right]^{1/2}
\nonumber\\
&&+ {1 \over 2} \left|\alpha^{\rm cr}  \, K_z\right|
- \etat \, K_z^2 ,
\label{D11}
\end{eqnarray}
where $\Omega_A=K_z \overline{V}_A$,
$\Omega^{\rm cr}=
c^{-1} \, \overline{J^{\rm cr}} \,(4\pi /\overline{\rho})^{1/2}$
and $\alpha^{\rm cr}=\alpha^{\rm cr}_{xx}
=\alpha^{\rm cr}_{yy}=\alpha^{\rm cr}_1 +
\alpha^{\rm cr}_2$.
The growth rate $\gamma_{\rm inst}$ of the dynamo
instability can be interpreted as the
interaction of the $\alpha^2$ large-scale dynamo instability
[determined by the terms $\propto \alpha^{\rm cr} K_z$
in \Eq{D11}] and the Bell instability
[described by the terms $\propto \left|\Omega_A \,
\Omega^{\rm cr}\right| -\Omega_A^2$ in \Eq{D11}]
in a homogeneous turbulent plasma with a cosmic rays current.
When $\alpha^{\rm cr} = 0$,
the growth rate of the large-scale
instability (\ref{D11}) coincides with that derived by
\cite{BOE10}; see Eq.~(23) of their paper.

\section{Direct numerical simulations}

\subsection{DNS model}

We consider a cubic computational domain of size $L^3$.
The smallest wavenumber is $k_1=2\pi/L$.
We adopt an isothermal equation of state with constant
sound speed $\cs$, so the gas pressure is $p=\rho\cs^2$.
We solve the equations of compressible MHD in the form
\begin{eqnarray}
\rho{\DD\UU\over\DD t}&=& {1 \over 4 \pi} (\nab\times\BB)\times\BB
-\cs^2\nab \rho+\nab\cdot(2\nu\rho\SSSS)
\nonumber\\
&& -{1 \over c}\JJ^{\rm cr}\times\BB ,
\label{AA1}\\
{\partial\AAA\over\partial t}&=& \UU\times\BB+\eta\nabla^2\AAA,
\label{AA2}\\
{\partial\rho\over\partial t}&=& -\nab\cdot\rho\UU,
\label{AA3}
\end{eqnarray}
where $\nu$ and $\eta$ are kinematic viscosity
and magnetic diffusivity,
respectively,  $\BB=\BB_0+\nab\times\AAA$
is the magnetic field consisting
of a uniform mean background field, $\BB_0=(0,0,B_0)$,
and a nonuniform part
that is represented in terms of the magnetic vector
potential $\AAA$,
and ${\sf S}_{ij} = \half(U_{i,j}+U_{j,i})
-\onethird \delta_{ij} \nab\cdot\UU$ is the traceless
rate of strain tensor, where commas denote
partial differentiation.

In all cases we adopt triply periodic boundary conditions.
The simulations are performed with the {\sc Pencil Code}%
\footnote{{\tt http://pencil-code.googlecode.com}},
which uses sixth-order explicit finite differences in space and a
third-order accurate time stepping method \citep{BD02}.
Simulations have been done with various resolutions, but here we
focus on two runs with $512^3$ mesh points.
Some of the results are also compared with
corresponding ones at $256^3$ mesh points.

\begin{figure*}
\begin{center}
\includegraphics[width=\textwidth]{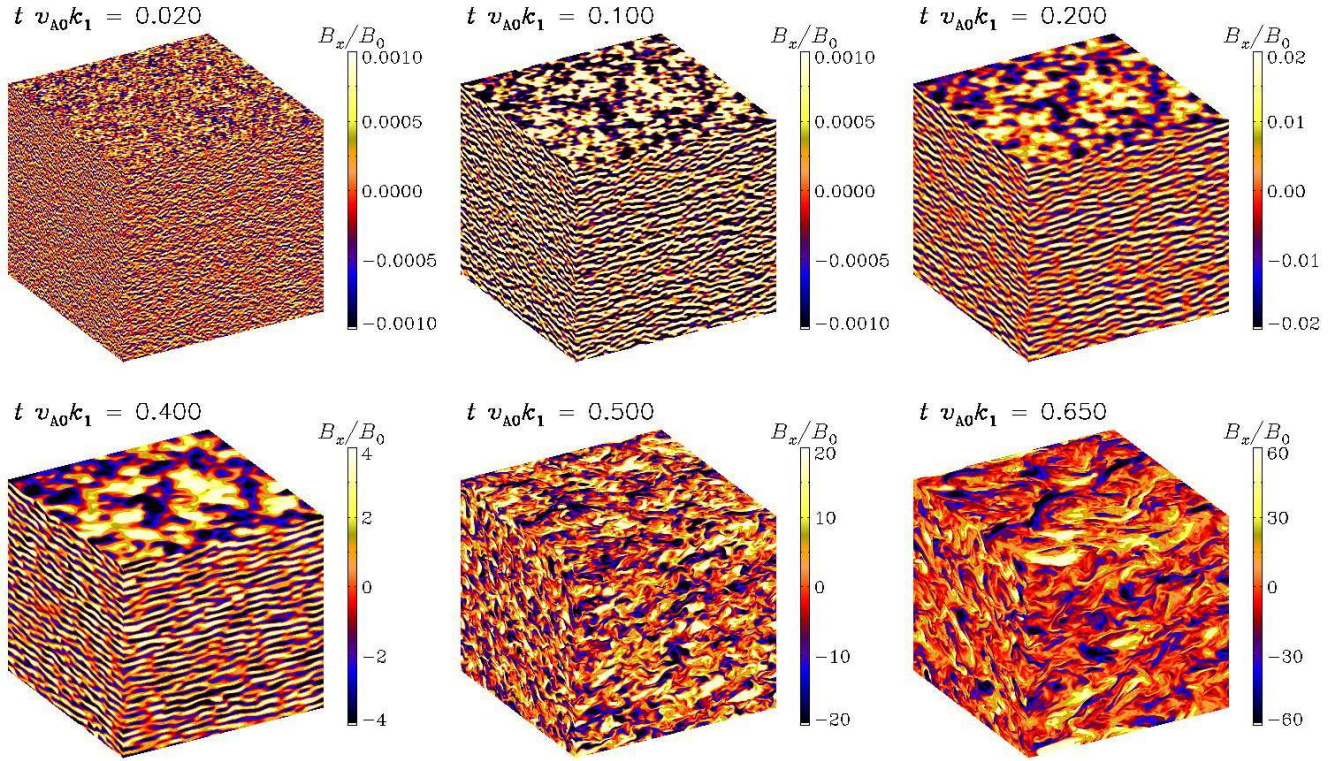}
%\plotone{f1}
\end{center}\caption[]{
Visualization of $B_x/B_0$ on the periphery of the
computational domain using $512^3$
mesh points for $J^{\rm cr}=0.1$, $B_0=0.01$, $k_1=1/8$
(so that ${\cal J}=80$), and $\nu=\eta=10^{-3}$
(so that the Lundquist number $\Lu=80$).
}\label{Bx_T512eo4x}
\end{figure*}

\subsection{Test-field method}
\label{Test_field_method}

We apply the quasi-kinematic test-field method
\citep[see, e.g.,][]{Sch05,Sch07,BRRS08} to determine
all relevant components of the tensor $\alpha_{ij}$
and turbulent magnetic diffusion.
This method allows for the presence of strong magnetic field
as long as the magnetic fluctuations are entirely a consequence
of the imposed field \citep{RB10}.
The essence of this method is that a set of prescribed
test fields $\meanBB^{(p,q)}$ and the flow from
the DNS are used to evolve separate realizations of
small-scale fields ${\bm b}^{(p,q)}$.
Neither the test fields $\meanBB^{(p,q)}$ nor the
small-scale fields ${\bm b}^{(p,q)}$
act back on the flow.
These small-scale fields are then used to
compute the electromotive force $\meanEMF^{(p,q)}$
corresponding to the test field $\meanBB^{(p,q)}$.
The number and form of the test
fields used depends on the problem at hand.

The choice of test fields depends on the averaging
that is performed.
Relevant for the present study are averages that depend
on $x$ or $y$, or both.
To gather sufficient statistics, we adopt planar $yz$ averages
that depend on $x$, so we use test fields
$\meanBB^{(1c)}=(0,\tilde B_0\cos kx,0)$,
$\meanBB^{(1s)}=(0,\tilde B_0\sin kx,0)$,
$\meanBB^{(2c)}=(0,0,\tilde B_0\cos kx)$, and
$\meanBB^{(2s)}=(0,0,\tilde B_0\sin kx)$,
in which case the series expansion of the electromotive
force contains two terms
\begin{equation}
\mathcal{E}_i=a_{ij}\mean{B}_j - \eta_{ij}(\nab\times\meanBB)_j.
\end{equation}
The symmetric part of the tensor $a_{ij}$ is of particular
interest and is commonly referred to as the $\alpha$ tensor,
\begin{eqnarray}
\alpha_{ij}=\half(a_{ij}+a_{ji}).
\end{eqnarray}
Errors are estimated by dividing the time series into three equally
long parts and computing time averages for each of them. The largest
departure from the time average computed over the entire time series
represents an estimate of the error.

\subsection{DNS results}
\label{DNS_results}

All runs are isothermal with $\cs=10$.
The box has a size of $16\pi$, i.e., $k_1=1/8$.
In our units, $4\pi/c=1$, $\rho_0=1$ and
the magnetic Prandtl number $\nu/\eta=1$.
The relevant non-dimensional parameters are
$\Lu=\vAz/\eta k_1$ (the Lundquist number),
where $\vAz=B_0$ is the non-dimensional Alfv\'en speed, and
${\cal J}=4\pi J^{\rm cr}/c B_0 k_1$ (the non-dimensional
cosmic ray current density).

\begin{figure}\begin{center}
\includegraphics[width=\columnwidth]{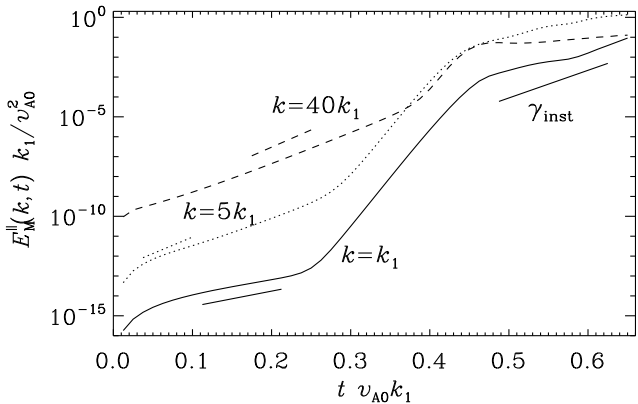}
%\plotone{f2}
\end{center}\caption[]{
Time evolution of $E_{\it M}^\parallel \, k_1/\vAz^2$ for modes with different
wavenumbers for the run with ${\cal J}=80$.
The short straight lines show the growth rate of the Bell instability,
as given by \Eq{A7} for modes with three selected values of $k$,
as well as the value of $\gamma_{\rm inst}$ as given by \Eq{D17}.
}\label{pspec_compt2_tot}
\end{figure}

Visualizations of the magnetic field $B_x/B_0$ on the periphery
of the computational domain are shown in \Fig{Bx_T512eo4x}
for a run with $512^3$ mesh points using the following parameters:
the non-dimensional cosmic ray current density is
${\cal J}=80$ and the Lundquist number $\Lu=80$
(with $J^{\rm cr}=0.1$, $B_0=0.01$ and $k_1=1/8$).
In the beginning of the instability, the length scale is rather
small, but it increases continuously as time goes on.
Note in particular the much larger horizontal length scales
in the $xy$ plane that may be associated with the dynamo instability.
After $t\vAz k_1\approx0.5$ the instability reaches yet another stage
during which the magnetic field pattern becomes turbulent.
Again, as time goes on the typical eddy scale increases.

\begin{figure}\begin{center}
\includegraphics[width=\columnwidth]{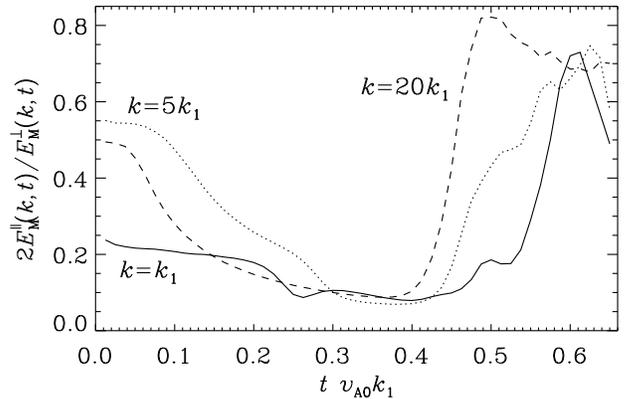}
%\plotone{f3}
\end{center}\caption[]{
Time evolution of the ratio of the spectral vertical (along the
imposed field ${\bm B}_0$) and horizontal magnetic energies $2
E_{\it M}^\parallel/E_{\it M}^\perp$ for the run with ${\cal J}=80$.
} \label{pspec_compt2_rat}
\end{figure}

\begin{figure}
\begin{center}
\includegraphics[width=\columnwidth]{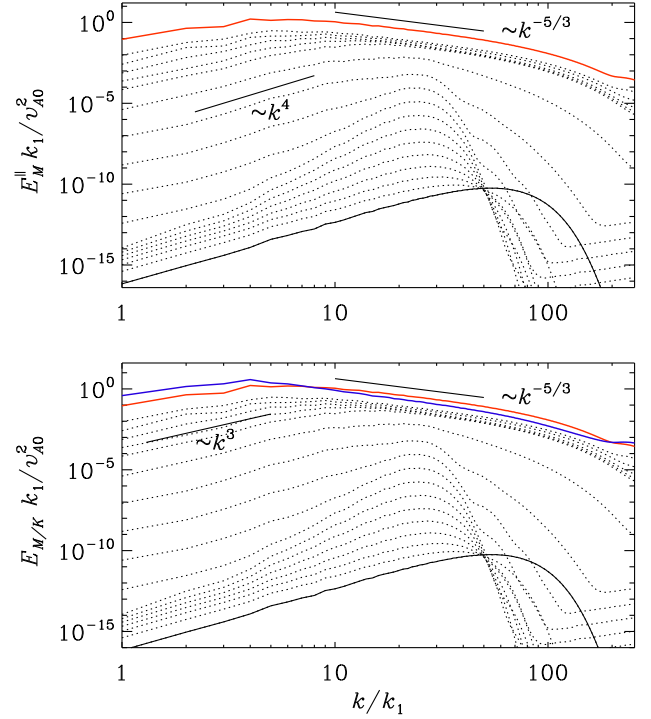}
%\plotone{f4}
\end{center}\caption[]{
Time evolution of $E_{\it M}^\parallel$ and $E_{\it M}$ for the run
using $512^3$ mesh points ($J^{\rm cr}=0.1$, $B_0=10^{-2}$, so that
${\cal J}=80$). The normalized time interval between different
spectra is $\vAz k_1\Delta t=3/80\approx0.04$. The solid lines refer
to the initial spectra proportional to $k^4$ for small values of $k$
and the red and blue lines represent the last instant of $E_{\it M}$
and $E_{\it K}$, respectively. The straight lines show the $k^4$ and
$k^{-5/3}$ power laws.
}\label{ppower_all_EK_EM_T512eo4x_powerbz}
\end{figure}

All runs with a constant cosmic ray current show that
there is a growth until the velocities become eventually supersonic.
This is probably the reason the simulation terminates.
It is conceivable that this could be avoided by including
the backreaction of the amplified field on the cosmic ray current,
which would limit the Bell instability \citep{Spitkovsky}.
In Figs.~\ref{pspec_compt2_tot} and \ref{pspec_compt2_rat}
we show the time evolution of the normalized spectral vertical
magnetic energies $E_{\it M}^\parallel \, k_1 / \vAz ^2$
and the ratio of the spectral vertical
(along the imposed field ${\bm B}_0$) and horizontal
magnetic energies $2E_{\it M}^\parallel/E_{\it M}^\perp$ for modes with
different wavenumbers for the same run as in \Fig{Bx_T512eo4x}.
As follows from these figures the dynamics of the instability
has the following stages:
\begin{itemize}
\item[(i)] In the early stage there is the development of small-scale
instability that results in the production of small-scale turbulence.
It is seen in \Fig{pspec_compt2_tot} that the mode-averaged growth rate of
this instability in this stage is slightly smaller than that of
\Eq{A7} that describes the Bell instability for the
fastest growing mode of a given $|k|$, as should be expected.
\item[(ii)] In the second stage, there is formation of large-scale
magnetic structures (during the interval $0.2$--$0.4$
Alfv\'en times; see \Fig{Bx_T512eo4x}).
Note from \Fig{pspec_compt2_tot} that the growth rate of the large-scale
mode $k=k_1$ is about twice the growth rate of the largest mode.
The interpretation of this will be given below.
\item[(iii)] In the final stage, there is a
development of larger-scale turbulence; the perturbed field actually
exceeds the original field by a considerable factor, and a significant
fraction of the energy is now present in modes with
$k_{\perp} \simeq k_{\parallel}$.
The growth rate in this final stage agrees with that predicted by \Eq{D17},
where $\alpha^{\rm cr}_{yy}$, $\alpha^{\rm cr}_{zz}$, and $\etat$ have
been obtained with the test-field method, as described below.
\end{itemize}
Time evolution of power spectra of magnetic energy of the $B_z$ component,
$E_{\it M}^\parallel$, and those of all components,
$E_{\it M}$, are shown in \Fig{ppower_all_EK_EM_T512eo4x_powerbz},
which demonstrates an inverse energy cascade-like behavior.
The evolution during the second stage (during the interval $0.2$--$0.4$
Alfv\'en times; see \Fig{Bx_T512eo4x}), shows that large scale modes
($k\sim k_1$) are growing at about {\it twice} the growth rate
of the fastest growing mode $k/k_1 \sim 40)$.
We interpret this as perturbation field growth $\partial \bb/\partial t$
at large scale due to the coupling of pairs of higher $k$ modes
($k_1$ and $k_2$, say), and therefore proportional to
$\urms^2 B_0 \propto \exp(\gamma_1 + \gamma_2)$, where $\gamma_1$ and
$\gamma_2$ are the respective linear growth rates of the high $k$ modes.
During this stage, the perturbed field is still small compared to the
original field, so the exponential growth of the total field as described
by equations~(\ref{D3}) and ~(\ref{D4}) has not yet begun.

\begin{figure}
\begin{center}
\includegraphics[width=\columnwidth]{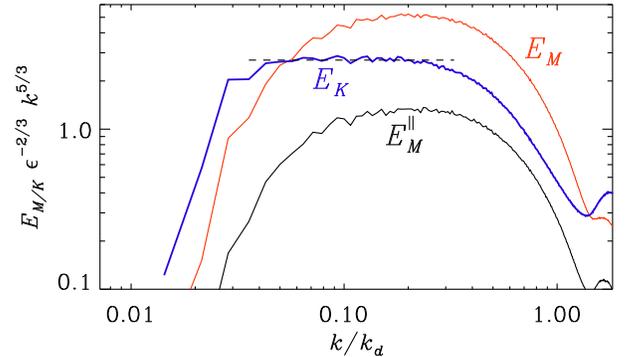}
%\plotone{f4b}
\end{center}\caption[]{
Compensated spectra of $E_{\it K}$ (blue), $E_{\it M}$ (red), and
$E_{\it M}^\parallel$ (black), at the end of the simulation.
Here, $\epsilon$ is the total (kinetic and magnetic) energy dissipation rate.
The dashed horizontal line goes through 2.7.
}\label{ppower_all_comp_EK_EM_T512eo4x_powerbz}
\end{figure}

At the end of the simulation, both kinetic and magnetic energy spectra
develop a $k^{-5/3}$ energy spectrum.
This is shown in \Fig{ppower_all_comp_EK_EM_T512eo4x_powerbz}, where
we plot spectra compensated by $\epsilon^{-2/3}k^{5/3}$, where
$\epsilon$ is the total (kinetic and magnetic) energy dissipation rate.
The wavenumber is normalized by the dissipation wavenumber
$k_d=[\epsilon/(\nu+\eta)^3]^{1/4}$.

\begin{figure}\begin{center}
\includegraphics[width=\columnwidth]{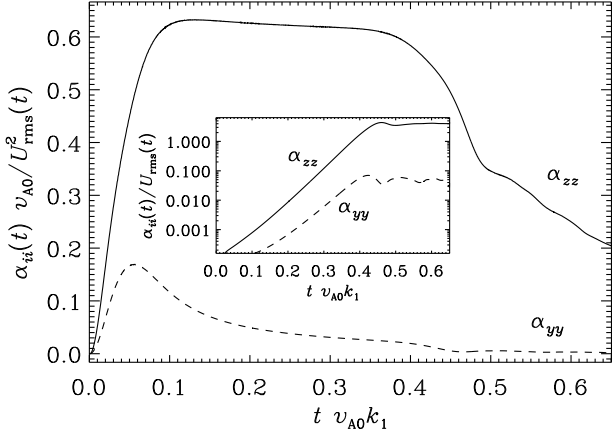}
%\plotone{f5}
\end{center}\caption[]{
Normalized $\alpha_{zz} \, \vAz / u_{\rm rms}^2$ and $\alpha_{yy} \,
\vAz / u_{\rm rms}^2$ for the run with ${\cal J}=80$. The inset
shows that after about 0.45 Alfv\'en times, both $\alpha_{yy}^{\rm
DNS}/\urms$ and $\alpha_{zz}^{\rm DNS}/\urms$ are approximately
constant in time. }\label{palptest_comp}
\end{figure}

\begin{figure}\begin{center}
\includegraphics[width=\columnwidth]{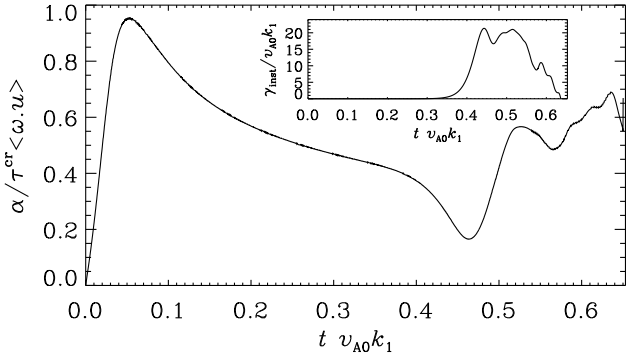}
%\plotone{f6}
\end{center}\caption[]{
Evolution of $\alpha=(\alpha_{yy}\alpha_{zz})^{1/2}$, normalized by
$\tau^{\rm cr}\langle\oo\cdot\uu\rangle$ where $\tau^{\rm
cr}=1/\omega^{\rm cr}$ for the run with ${\cal J}=80$.
The inset gives the instantaneous value of $\gamma_{\rm inst}$
as derived from \Eq{D17}.
}\label{palp}\end{figure}

The test-field results for the normalized components of the $\alpha$ tensor,
$\alpha_{zz} \, \vAz / u_{\rm rms}^2$ and $\alpha_{yy} \, \vAz / u_{\rm rms}^2$
are shown in \Fig{palptest_comp} for the same run.
The value of $\alpha_{zz}^{\rm DNS} \, \vAz / u_{\rm rms}^2 \approx 0.6$
is of the same order of magnitude as that determined by
\Eq{D4}, i.e., $\alpha_{zz}^{\rm theory} \, \vAz
/ u_{\rm rms}^2 \approx 0.5$.
Note that at late times, after about 0.45 Alfv\'en times, we find that
both $\alpha_{yy}^{\rm DNS}/\urms$ and $\alpha_{zz}^{\rm DNS}/\urms$
are approximately constant in time.
This measured value of the $\alpha$ effect is much larger
than that based on kinetic helicity determined by \Eq{D3},
unless $\tau_0$ is considerably larger than $1/k\urms$ see \Fig{palp}.
We interpret this as evidence for an additional contribution, \Eq{D4}.
The growth rate of the large-scale dynamo instability is of the same
order as that theoretically predicted from the large-scale dynamo
instability.
Indeed, \Eq{D17} for the growth rate yields
$\gamma^{\rm theory}\approx 0.03$, corresponding to
$\gamma^{\rm theory}/\vAz k_1\approx24$ in non-dimensional units.
This value is in agreement with the growth rate expected from an
$\alpha^2$ dynamo with coefficients obtained with the test-field method;
see the inset of \Fig{palp}, where $\gamma_{\rm inst}/\vAz k_1\approx20$.
This agrees with the growth seen in the DNS at $k=k_1$ during the time
interval $0.45$--$0.65$; see \Fig{pspec_compt2_tot}.
Finally, \Eq{D18} for the ratio of magnetic energies
along and perpendicular to the direction of the cosmic ray current yields
$\big[\tilde{B}_z^2/(\tilde{B}_x^2+\tilde{B}_y^2)\big]^{\rm theory}\approx0.08$,
while this ratio according to DNS (see \Fig{pspec_compt2_rat}) is estimated as
$\big[\tilde{B}_z^2/(\tilde{B}_x^2+\tilde{B}_y^2)\big]^{\rm DNS}\approx0.06$.
The above comparison implies a good agreement between theoretical
predictions and the results of the numerical simulations.

\begin{figure}
\begin{center}
\includegraphics[width=\columnwidth]{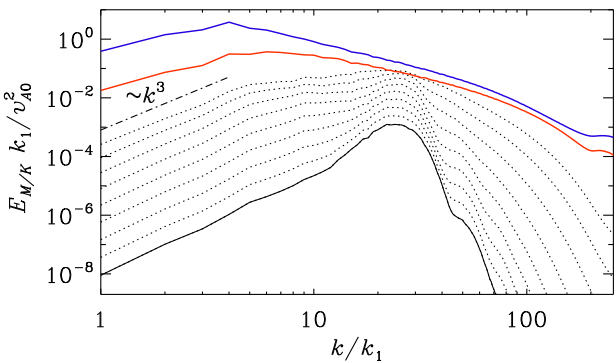}
%\plotone{f7}
\end{center}\caption[]{
Same as the bottom panel of
\Fig{ppower_all_EK_EM_T512eo4x_powerbz} (${\cal J}=80$), but
showing only the time interval $0.35\leq t\vAz k_1\leq 0.45$ (dashed
lines), i.e., the last 0.1 Alfv\'en times just near the end of the
linear growth phase. The black solid lines refer to $t\vAz
k_1=0.35$, while red lines and blue lines refer to $\kf^{\it M}(t)$
and $\kf^{\it K}(t)$ at $t\vAz k_1=0.65$.
The slope $k^3$ is shown for comparison.
}\label{ppower_last_EK_EM_T256eo4x_powerbz}
\end{figure}

In \Fig{ppower_last_EK_EM_T256eo4x_powerbz} we show the time evolution
of power spectra of magnetic energy of the $B_z$ component,
$E_{\it M}^\parallel$, and those of all components, $E_{\it M}$, like it was
presented in \Fig{ppower_all_EK_EM_T512eo4x_powerbz}, but during
only the last $1/10$ of an Alfv\'en time. We see that the amplification
of magnetic field with respect to initial perturbations
during the last $1/10$ of an Alfv\'en time (which is of the order of the
shock crossing time) at $k=k_1$ is $3.5$ orders of magnitude
(i.e., up to $0.6 \times 10^{-3}$ of the field).
However this field is less than the equilibrium imposed field
$B_0=10^{-2}$.
On the other hand, the visualizations of the magnetic field $B_x/B_0$
on the periphery of the computational domain shown in \Fig{Bx_T512eo4x},
demonstrates that the ratio $B_x/B_0 < 60$ at $t\vAz k_1=0.65$.

\begin{figure}
\begin{center}
\includegraphics[width=\columnwidth]{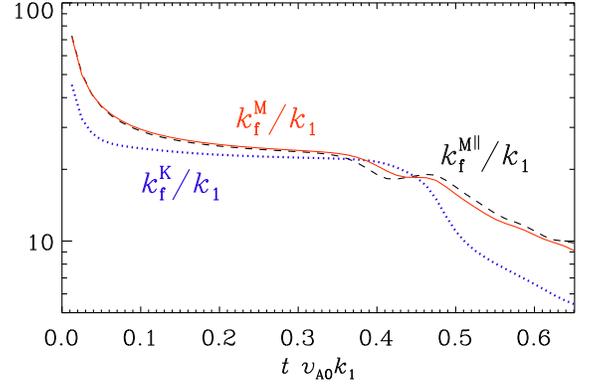}
%\plotone{f8}
\end{center}\caption[]{
Evolution of $\kf^{\it M}$ (solid, red), $\kf^{M\parallel}$ (dashed),
and $\kf^{\it K}$ (dotted, blue)
for the same run as in \Fig{Bx_T512eo4x} (${\cal J}=80$).
}\label{ppower_kf_T512eo4x_powerbz}
\end{figure}

To quantify the inverse cascade-like behavior, let us now look at
the evolution of the wavenumber of the magnetic energy-carrying eddies,
$\kf^{\it M}$, defined via
\begin{eqnarray}
\left[\kf^{\rm M}(t)\right]^{-1}
\!=\!\left.\int k^{-1} E_{\rm M}(k,t)\,\dd k\right/
\!\!\int E_{\rm M}(k,t)\,\dd k,
\label{D227}
\end{eqnarray}
and likewise for kinetic energy as well as magnetic energy in the
$z$ component, $\kf^{\it K}$ and $\kf^{\it M\parallel}$, respectively
(see \Fig{ppower_kf_T512eo4x_powerbz}).
It turns out that all three wavenumbers reach a value somewhat above
$20\,k_1$ by the end of the small-scale dynamo instability,
and then drop rapidly in the mean-field dynamo stage.
Note, however, that the decrease of $\kf^{\it K}$ is somewhat faster
than that of $\kf^{\it M}$.

\begin{figure}
\begin{center}
\includegraphics[width=\columnwidth]{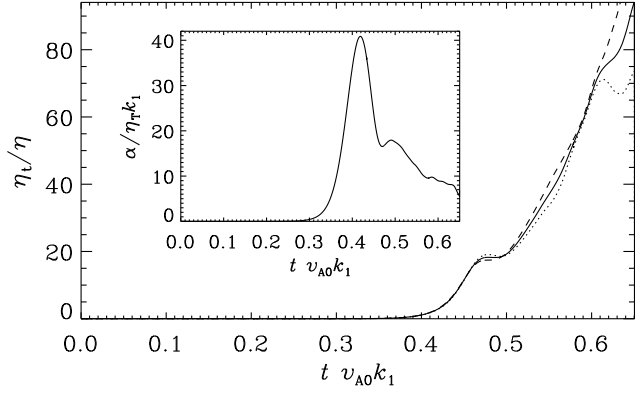}
%\plotone{f9}
\end{center}\caption[]{
Evolution of $\etat/\eta$ (solid line), $\eta_{yy}/\eta$ (dotted
line), and $\eta_{zz}/\eta$ (dashed line) for the run ${\cal J}=80$.
The inset shows $\alpha/\etat k_1$. }\label{peta}
\end{figure}

The test-field method yields not only $\alpha_{ij}$, but also the components
of the turbulent magnetic diffusivity tensor $\eta_{ij}$.
It turns out that $\eta_{yy}\approx\eta_{zz}$.
The ratio $\etat/\eta$, where $\etat=(\eta_{yy}+\eta_{zz})/2$,
greatly exceeds unity in the late stages; see \Fig{peta}.
Mean-field dynamo efficiency depends on the dynamo number
$D=(\alpha/\etaT k_1)^2$,
where $\alpha=(\alpha_{yy}\alpha_{zz})^{1/2}$ and $\etaT=\etat+\eta$.
The dynamo number exceeds the critical value for dynamo action of unity
$(D>1)$ after about 0.3 Alfv\'en times; see the inset of \Fig{peta}.

\subsection{Comparison with the run ${\cal J}=800$}

\begin{figure}
\begin{center}
\includegraphics[width=\columnwidth]{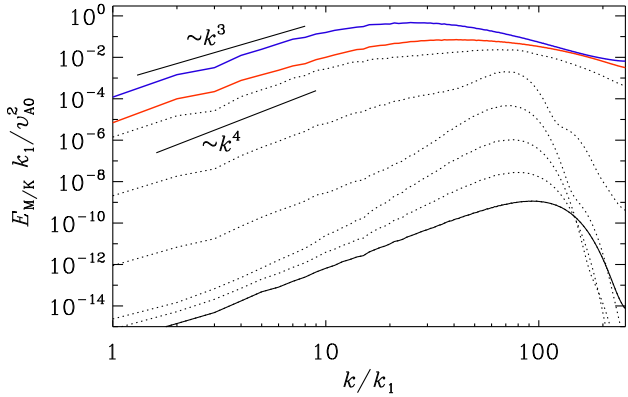}
%\plotone{f10}
\end{center}\caption[]{
Time evolution of $E_{\it M}$ and $E_{\it K}$ for the run ${\cal
J}=800$ ($J^{\rm cr}=1$, $B_0=10^{-2}$) showing only the time
interval $0.0125 \leq t\vAz k_1\leq 0.075$. The time interval
between the lines $\Delta t \vAz k_1=0.0125$. The red and blue lines
represent the last instant of $E_{\it M}$ and $E_{\it K}$,
respectively.
The slopes $k^4$ and $k^3$ are shown for comparison.
}\label{ppower_all_EK_EM_T512eo5x_powerbz}
\end{figure}

In this subsection we discuss the results of
DNS with the higher cosmic ray current, i.e.,
when the normalized cosmic ray current is
increased by one order of magnitude ${\cal J}=800$
(i.e., $J^{\rm cr}=1$, $B_0=0.01$ and
$k_1=1/8$), and the other parameters are the same as
in the previous subsection (the Lundquist number
$\Lu=80$ and the resolution is $512^3$ mesh points).
In this case very small
scales (higher $k$ modes) are not well resolved.
On the other hand, the main contribution to the
mean-field dynamo (which is the main subject of
our study) is of the maximum scale of turbulent motions.
The time evolution of spectra of
magnetic, $E_{\it M}$, and kinetic, $E_{\it K}$,
energies for ${\cal J}=800$ is shown in
\Fig{ppower_all_EK_EM_T512eo5x_powerbz}, that
demonstrate an inverse energy cascade-like
behavior, but to a lesser extent in comparison with
the lower CR current (${\cal J}=80$).

\begin{figure}
\begin{center}
\includegraphics[width=\columnwidth]{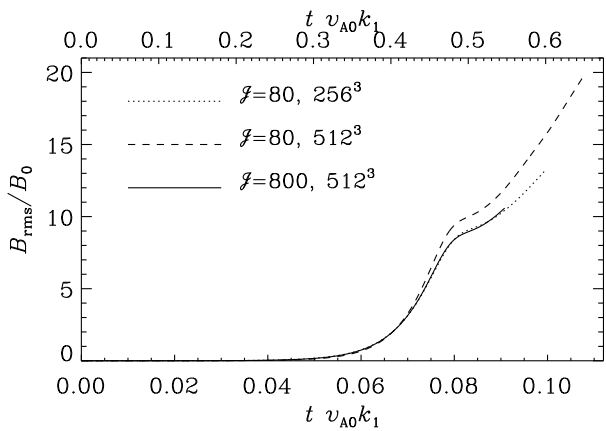}
%\plotone{f11}
\end{center}\caption[]{
Normalized $B_{\rm rms}/B_0$ for runs with ${\cal J}=80$
(using $256^3$ and $512^3$ meshpoints; dotted and dashed lines,
respectively) and ${\cal J}=800$ ($512^3$ meshpoints, solid
line). Upper horizontal axis corresponds to case
${\cal J}=80$, while lower horizontal axis
corresponds to case ${\cal J}=800$. }
\label{pcomp_BB0}
\end{figure}

\begin{figure}
\begin{center}
\includegraphics[width=\columnwidth]{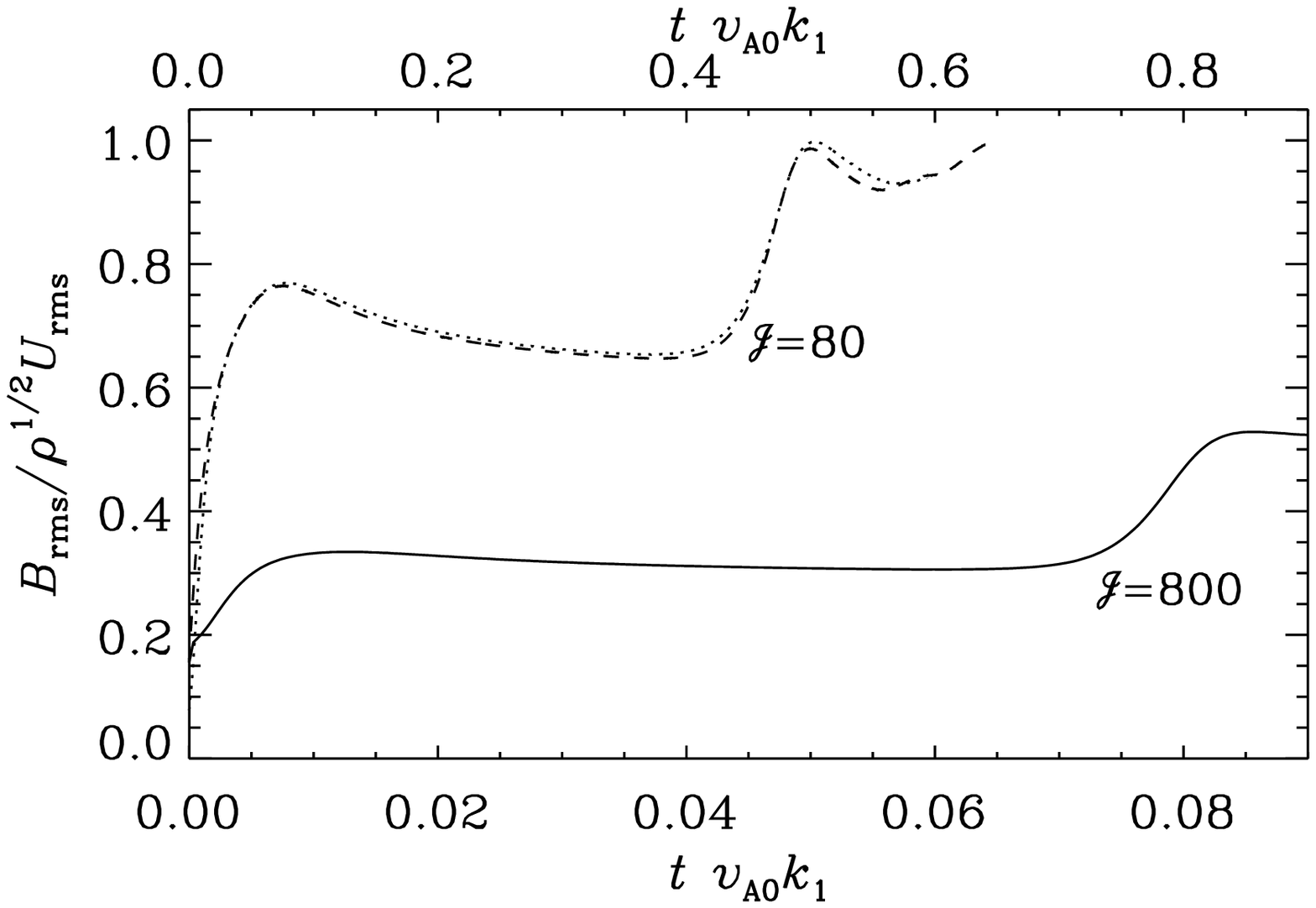}
%\plotone{f12}
\end{center}\caption[]{
Evolution of $B_{\rm rms}/\rho^{1/2} U_{\rm rms}$
for runs with ${\cal J}=80$ (dotted and dashed lines for
$256^3$ and $512^3$ meshpoints, respectively) and
${\cal J}=800$ (solid line, $512^3$ meshpoints). Upper horizontal
axis corresponds to case ${\cal J}=80$, while
lower horizontal axis corresponds to case ${\cal
J}=800$. }\label{pcomp_BU}
\end{figure}

\begin{figure}
\begin{center}
\includegraphics[width=\columnwidth]{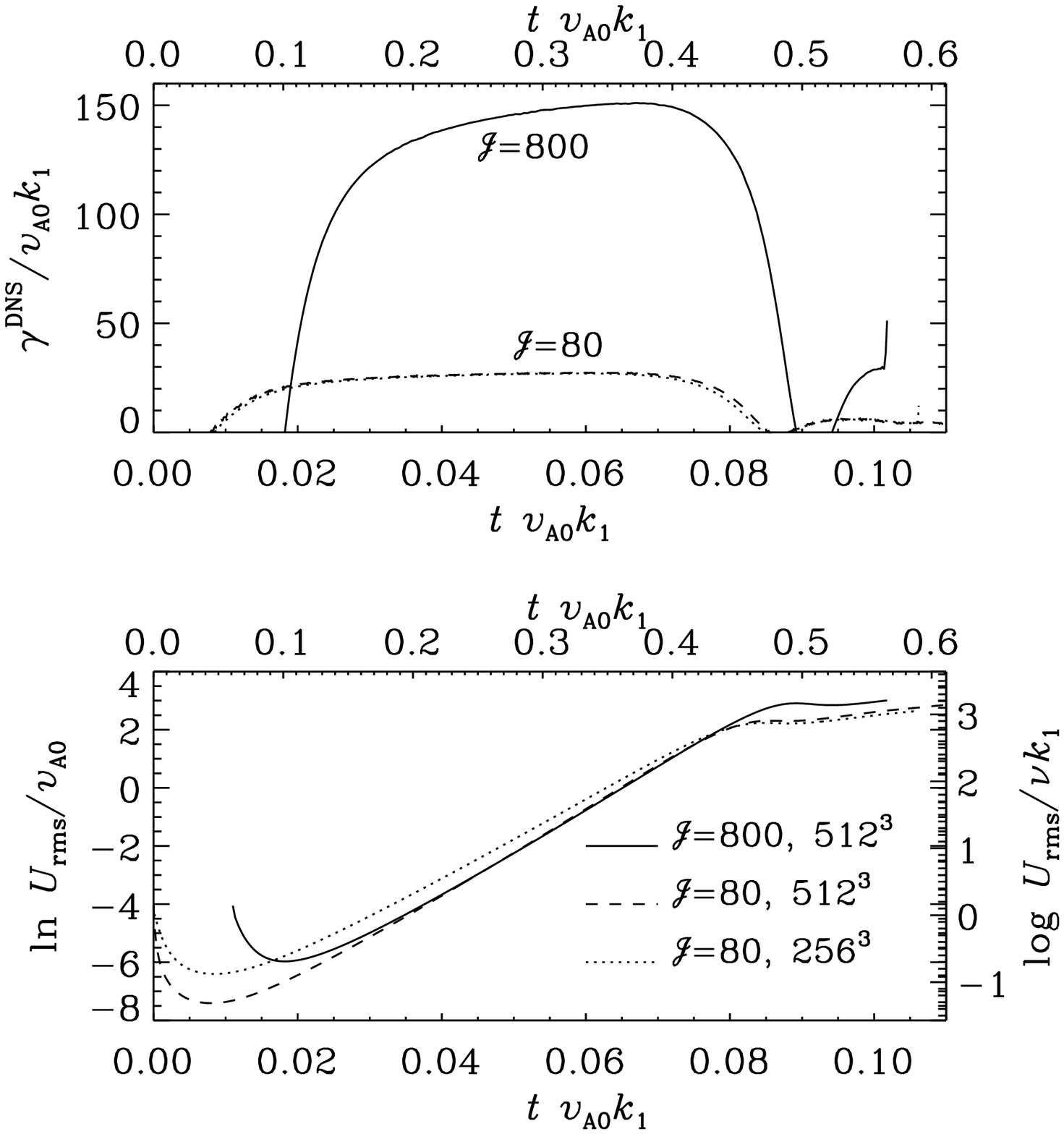}
%\plotone{f13}
\end{center}\caption[]{
Instantaneous growth rate $\gamma^{\rm DNS}=\dd\ln U_{\rm rms}/\dd t$
(upper panel) and $\ln U_{\rm rms} /\nu k_1$ (lower panel) for
runs with ${\cal J}=80$ (dotted and dashed lines for
$256^3$ and $512^3$ meshpoints, respectively) and ${\cal
J}=800$ (solid line, $512^3$ meshpoints). Upper horizontal axis
corresponds to case ${\cal J}=80$, while lower
horizontal axis corresponds to case ${\cal
J}=800$.} \label{pcomp_lam}
\end{figure}

For comparison with the case ${\cal J}=80$,
in Figs.~\ref{pcomp_BB0}--\ref{pcomp_lam} we show
the time evolution of the total magnetic field
$B_{\rm rms}/B_0$, $B_{\rm rms}/\rho^{1/2} U_{\rm
rms}$ (which does not include the imposed field
$B_0$) and the growth rate of the total velocity
field, $\gamma^{\rm DNS}=\dd\ln U_{\rm rms}/\dd
t$, for runs with ${\cal J}=80$ (dotted and dashed lines
for $256^3$ and $512^3$ meshpoints, respectively) and
${\cal J}=800$ (solid line).
Since the time evolution for different values of the cosmic ray
current occurs on different time scales, we use
the upper horizontal axis for the case ${\cal J}=80$,
while the lower horizontal axis
corresponds to case ${\cal J}=800$.

\begin{figure}
\begin{center}
\includegraphics[width=\columnwidth]{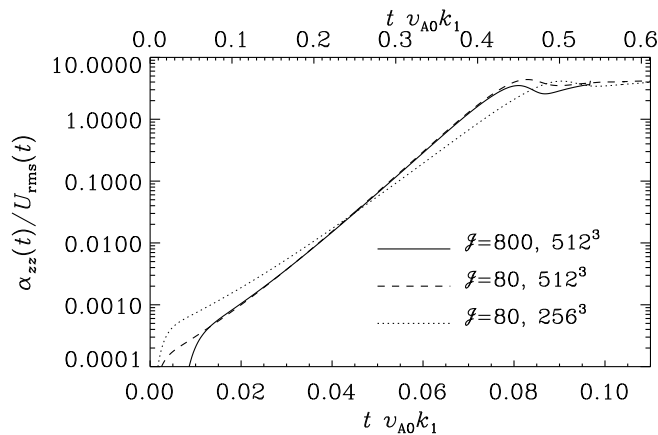}
%\plotone{f14}
\end{center}\caption[]{
Normalized $\alpha_{zz} / U_{\rm rms}$ for runs with ${\cal J}=80$
(dotted and dashed lines for $256^3$ and $512^3$ meshpoints, respectively)
and ${\cal J}=800$ (solid line, $512^3$ meshpoints). Upper horizontal axis
corresponds to case ${\cal J}=80$, while lower horizontal axis
corresponds to case ${\cal J}=800$. } \label{pcomp_alp}
\end{figure}

Inspection of \Fig{pcomp_BB0} shows that, at the
final stage of evolution, the generated magnetic
field is by one order of magnitude larger than
the imposed field $B_0$.
The ${\cal J}=80$ results at lower resolution are similar to
those at higher, but have run for a slightly shorter time
before resolution problems occurred.
The growth rate of the velocity and magnetic field is 5
times larger for the case of ${\cal J}=800$ (see
the upper panel of \Fig{pcomp_lam}).
On the other hand, the evolution of the kinetic energy -- apart
from this rescaling of time -- is not
strongly dependent on the cosmic ray current (see
the lower panel of \Fig{pcomp_lam}), and neither is
$\alpha_{zz} / U_{\rm rms}$ (see the test-field
results of the measured $\alpha$ effect in
\Fig{pcomp_alp} shown in log scale).

As noted above, the quasi-kinematic test-field method is valid as long as
the magnetic fluctuations are entirely a consequence of the imposed field.
In one particular case we have verified this by comparing with results from
a fully nonlinear test-field method where velocity fluctuations resulting
from the interaction with magnetic field fluctuations are also included.
This method has currently been tested and implemented
in the {\sc Pencil Code} only for a modified set of
equations in which the pressure gradient and the $\UU\cdot\nab\UU$ term
are omitted, but the Lorentz force is fully retained \citep{RB10}.
We have applied this method, with the modified set of equations, to a case
similar to that displayed in \Fig{palptest_comp}, but at lower resolution
($64^3$).
In that case, $\alpha_{yy}\vA/\Urms^2$ is nearly constant after
$t\vA k_1=0.1$ and comparable to the corresponding value shown in
\Fig{palptest_comp} at $t\vA k_1=0.4$, while $\alpha_{zz}\vA/\Urms^2$
agrees with that of \Fig{palptest_comp} in the full time interval.
More importantly, however, the quasi-kinematic and fully nonlinear
test-field methods agree with each other within machine precision,
confirming thus the applicability of the quasi-kinematic method to the
present case.

\subsection{Interpretation of the Results}

Our results seem consistent with the following interpretation.
There appear to be three distinct stages.
In the first stage, the $ c^{-1} \, \JJ^{\rm cr}
\times {\bm b}$ force (which we refer to as the
Lorentz force due to the ``counter-CR current" --
i.e.\ the current in the thermal plasma that
cancels the cosmic ray current) amplifies the
motion perpendicular to the original, unperturbed
magnetic field in one circular polarization of
Alfv\'en modes, thereby stretching the field such
that it develops a component that is
perpendicular to the original direction.
This is the Bell instability in our simulation, but
anything that creates a perpendicular component
might work just as well for this phase. The Bell
instability grows fastest on small scales and in
a direction whose $\kk$ vector is parallel to the $z$ axis.
Tentatively, we may attribute the increasing
preference for ``perpendicular'' energy that we see
in the simulation, i.e.\ energy in motion in the
``perpendicular" ($x$ and $y$) directions, to the
faster growth rate of on-axis ($k=k_z$) Alfv\'en
waves. However, there is clearly significant
parallel energy, even in the linear growth
regime, presumably due to off-axis waves. The
reason that the ratio of ``parallel" to
``perpendicular" energy decreases in the first
stage is presumably because the on-axis waves,
which have only motion perpendicular to the axis,
are the fastest growing modes.

During the first stage, the ratio of
$\alpha_{zz}$ to $u_{\rm rms}$ grows exponentially,
because for any given mode, $\alpha \propto
\uu\cdot\nab\times\uu \sim k\uu^2$ is
proportional to $\uu^2$, and $\uu$ is growing
exponentially (\Fig{pcomp_lam}), consistent with
the linear Bell instability. The correlation
length of both the magnetic field and the
turbulence, apart from a drop in the very early
stages, remains more or less constant, and corresponds to the
scale of the fastest growing modes.

The second stage is much like the first, except that the growth rate
of the larger scale, low $k$, modes begins to increase above its linear
value; see \Fig{pspec_compt2_tot}.
This is presumably due to the fact that their growth is dominated
by nonlinear coupling of higher $k$ modes, which have developed much
larger amplitudes than the low $k$ ones, even though they are still in
the linear regime.

The third stage, for ${\cal J} = 80$ (${\cal J} = 800$), begins at
about $t\vAz k_1\sim 0.45$
($t\vAz k_1\sim 0.06$), as shown in the upper panel of \Fig{pcomp_lam}.
Several things clearly happen at the onset of the second stage: (i) The
growth of $U_{\rm rms}$ suddenly slows (\Fig{pcomp_lam}), (ii) the growth of the
magnetic field also shows a change (\Fig{pcomp_BB0}), though it continues to
rise, (iii) the {\it correlation lengths} of both the magnetic field and
the velocity begin to increase noticeably (\Fig{ppower_kf_T512eo4x_powerbz}),
(iv) the ratio of
$\alpha$ to $u_{\rm rms}$ stops rising exponentially and either flattens out
or grows much more slowly (Figs.~\ref{palptest_comp} and~\ref{pcomp_alp}),
(v) the turbulent magnetic diffusivity
begins to rise significantly and dominates the microscopic value
(\Fig{peta}), (vi) the ratio of energy in parallel magnetic field to that
in perpendicular field  rises sharply (\Fig{pspec_compt2_rat}),
and (vii) a Kolmogorov-type
spectrum is reached from below, while the level of turbulence
grows more slowly.

All these changes can be understood in terms of nonlinear effects.
Once the turbulent velocity exceeds the Alfv\'en velocity, the nonlinear
convective term in the Navier-Stokes equation becomes as or more important
than the Lorentz force due to the counter-CR current, so that the stirring
of the fluid by the latter, as expressed in unstable Alfv\'en modes, is in
equilibrium with eddy viscosity.
By the same token, the amplitude of the magnetic field is large, so
that the $\alpha^2$ dynamo is activated.
Taken in isolation, an exponentially growing $\alpha$ effect would
lead to super-exponential growth, but this is not what is seen.
The sudden rise of parallel to perpendicular magnetic
energy could perhaps be attributed to this effect,
but, looking at the shape of the spectra in
\Figs{ppower_all_EK_EM_T512eo4x_powerbz}{ppower_all_EK_EM_T512eo5x_powerbz},
and as said before, nonlinear mode coupling to the fastest growing modes,
which here turn out to be at $k/k_1\approx25$ and 70, respectively,
is a likely explanation.

In this {\it nonlinear stage}, the ratio of parallel to
perpendicular energy begins to grow, and the
field attains larger scale and becomes more isotropic, as
can be seen in \Fig{Bx_T512eo4x}.
The $\alpha^2$ dynamo can be
interpreted as an inverse cascade, in which
parallel and anti-parallel magnetic flux is
generated by the stretching of perpendicular
flux, while the anti-parallel flux is kinematically
``pumped" out of any given finite region of size
$L$ at a velocity of order $\alpha$ by the
$\alpha$ effect. The pumping of flux into large
regions of size $L$ can be viewed from a modal
point of view as inverse cascading of energy into
small wavenumbers of order $\pi/L$.

The observed scaling of the time evolution with
${\cal J}$ is reasonable: the timescale for
${\cal J} = 800$ is about a
factor of $5.5 \sim 10^{0.75}$ times less than
that for ${\cal J} = 80$. Because the stirring
force of the counter-CR current is proportional
to ${\cal J}$, one might expect velocities to
scale as ${\cal J}$ for a fixed correlation time.
However, if an acceleration $a$ operates over a
set distance $s$, then the velocity scales only
as $(s a)^{1/2}$. Once the turbulent velocity
exceeds $\vAz$, the amplitude of the transverse
motion associated with a given mode is $\sim
2\pi/k$, and the correlation length varies more
weakly than the correlation time with ${\cal J}$.
So we expect that the timescale for the dynamo
mechanism scales as ${\cal J}^{\delta}$, where
$0.5 \leq \delta \leq 1$, and this is what we
observe; $\delta \simeq 0.75$.  We have not
verified how this scaling law extends to larger
${\cal J}$.

In Kolmogorov turbulence, the
turbulent kinetic energy $\int E_{\it K}(k)\,\dd k$ at any
instant is proportional to $P^{2/3}$, where $P$
is the stirring power. In a situation such as the
present one, where the stirring force $F$ is
proportional to ${\cal J} B$, for a given
stirring scale $k$, the velocity $U$ scales as
$(F/k)^{1/2}$,  the power is proportional to
$F^{3/2}$ and the energy should then, by
dimensional arguments, have a finite, steady
state value that is  proportional to $F$. The
point is that for a given $B$, the energy $E$ has
a finite value to which it should rise and
flatten out. There is indication of this in our
simulation results, as seen in
Figs.~\ref{pspec_compt2_tot} and
\ref{pspec_compt2_rat}, where the total turbulent
energy flattens out at $t \vAz k_1 = 0.45$ for
${\cal J}= 80$ ($t \vAz k_1 = 0.08$ for ${\cal J}= 800$).
It does not completely flatten out
though, and we attribute this to the fact that
$B_{\rm rms}$ is still creeping up with time due
to the dynamo effect. Our simulations are not
long enough to determine whether  the magnetic
energy $E_M$ always reaches equipartition with
the kinetic energy $E_K$. If it does, then,
because $B$ scales as $E_M^{1/2}$, dimensional
arguments suggest that the force $F$, which
scales as ${\cal J} B$, therefore scales as
${\cal J} E_K^{1/2}$.
So, if $U$ scales as $(F/k)^{1/2}$ and thus $E_K$ scales as $F$,
then the force scales as ${\cal J}^2$. We believe it
would take longer simulation runs that were
presently feasible to check this prediction, and
whether $E_M$ indeed scales as $E_K$.

In the above discussion we have invoked the Bell
instability to generate magnetic flux that is
perpendicular to the local background, because
this, from the point of view of the fluid, is an
MHD effect and can be represented in an MHD
simulation.
We note, on the other hand, that non-MHD effects
could achieve the same result.
For example, the non-resonant firehose instability
could achieve the same stretching.
The helicity that is required
for the $\alpha$ effect relies on preferential
growth of one circular polarization over the
other, so the firehose instability on large
scales, which has no such preference, could not
by itself bring about an $\alpha$ effect.
However, it could combine with the resonant
cyclotron instabilities to do so. In this case,
the firehose instability would play the role that
differential rotation plays in the $\alpha
\Omega$ dynamo -- that of creating a perpendicular
field component, while the helical turbulence
that is generated by the Bell
instability would play the role of helical
turbulence that, in the  $\alpha \Omega$ dynamo,
is generated by the combination of convection in
a stratified medium and Coriolis force.

\section{Astrophysical Applications}

In a real astrophysical system, there is a
limited amount of time available. In the case of
an expanding blast wave, this is of order the
expansion time. Similarly, in an accretion shock,
accreting matter continuously sweeps magnetic
field downward over the crossing timescale of the
accreting matter.
The question is whether this is enough for significant magnetic
field amplification.

\subsection{Blast waves}

In our simulations, the growth   up to maximum
amplitude takes on the order of $0.45$ (0.08)
Alfv\'en crossing times across  the box for
${\cal J} = 80$ (${\cal J}  = 800$); see
Figs.~\ref{ppower_all_EK_EM_T512eo5x_powerbz}--\ref{pcomp_lam}.
However, this total time interval witnesses a
gain of many orders of magnitude
of the level of large-scale magnetic field,
because the simulation begins at the noise level.
For ${\cal J} = 80$ it can be seen from
Figs.~\ref{pspec_compt2_tot}
and~\ref{ppower_all_EK_EM_T512eo4x_powerbz} that
at $\vAz k_1 t \sim 0.3$, the growth at large
scales speeds up, the energy in field
perturbations at any given scale grows by a
factor of $\sim 10^{3/2}$ per $ \Delta t \sim
0.04 \vAz k_1$, and, at somewhat later times, by
a factor of nearly $10^2$ per $ \Delta t \sim
0.04 \vAz k_1$, suggesting that, during a period
of exponential growth, the gain factor over
an interval $T$ at wavenumber $k$, is $G(T,k) =
10^{2 v_A k_1 T/0.04}$.
For a supernova remnant of radius $R$, the precursor has a width $W$
of somewhat less than $R$, and the largest mode has
a wavenumber of order $k_1 = 2\pi /W \gtrsim 1$.
Thus, the available time $T$ for field
amplification by cosmic ray current is $T  = W/
u_s\sim 2\pi /(k_1 \vA M_A)$, where $M_A=u_s/\vA$ is the
Alfv\'en Mach number of the blast wave.
Then $\vA k_1 T \sim 2 \pi/M_A$,
suggesting that the $\alpha$ dynamo effect could
amplify the field energy of the remnant by a gain
factor $G$ of order $G = 10^{ \pi / 0.01 M_A}$,
and the amplification of the field's
magnitude would be the square root of this
factor, i.e. a factor of several for $M_A\sim 10^2$.
This appears to be consistent with \Fig{pcomp_BU}.
According to \Eq{KeyParameter}, and assuming $P^{\rm cr}\propto u_s^2$,
we see that raising the potential CR current by $\zeta^3$, speeds up
the time evolution of the field amplification by
$\zeta^{\delta}$, where $\delta$ is between 0.5 and 1;
see the discussion above.
So, very young SNR, where ${\cal J}$ can be much higher
than in the runs we made, the gain factor for an
expansion time could be higher.
In \Fig{ppower_all_EK_EM_T512eo4x_powerbz}, for
example, the magnetic energy on any given scale,
$[E_{\it K}(k)k]/(k_1/k)$ reaches, but does not
significantly exceed, unity.

For young, expanding supernova remnants $M_A$ is
typically $10^2$, so the growth factor on the
scale of the supernova could be of order one
$e$-fold per expansion time if the assumptions of
the simulation were to remain valid over the full
interval. While not dramatic, neither is it
insignificant in view of the uncertainties, a
modest amount of magnetic field amplification may
take place within an expanding supernova remnant,
and this may be enough to be compatible with
observational inferences \citep{Pohl05}.
While the correlation length of the magnetic
field increases with time, it is at all times
less than the size of the box by a factor of several.

On the other hand, increasing the shock velocity from $10^{-2}c$ to
near $c$ would raise ${\cal J}$ by a factor of $10^6$, thereby speeding
up the rate of field amplification by ${\cal J}^{6\delta} \ge 10^3$
while decreasing the expansion time by only a factor of $10^2$, so
there would then seem to be enough time to amplify the magnetic field
by many orders of magnitude even on the largest scale -- as demonstrated
in our simulations -- if there are no other fundamental limitations
that we have not yet identified.
Thus, GRBs, which create ultrarelativistic shocks in the
interstellar medium, would have the greatest potential for magnetic
field amplification, because their ratio of particle pressure to initial
magnetic pressure is so large.
This is now discussed below.

In our simulations, the velocity of the turbulence
attains a magnitude of about $10^2 \vAz$,   the
magnetic field increases to $\sim 10 $ times the
original field $B_0$, and the correlation length $L$ of
this field is about $10^{-1}$ of the box size $R$.
Thus the quantity $e B L$ is not much
changed from the original value $e B_0 R$,
meaning that the maximum energy attainable by
cosmic rays, $\sim u_s e B L/c$
\citep{Eichler-and-Pohl-2011},
is not much changed by the magnetic field amplification.
(However, until we are certain how the final correlation length scales with the
running time of the simulation, this matter remains not completely settled.)

If the energy of the highest
energy cosmic rays $E$ is taken to be $u_s B_0
R/c$ then the deflection of these highest energy
implied by (though ignored in the simulations) is
probably small for $u_{\rm rms}\ll u_s$.
In general, the turbulent rms velocity $u_{\rm rms}$ should
be somewhat less than the shock velocity, $u_s$,
so the potential drop $e u_{\rm rms} B L/c$
across one correlation length $L$ is somewhat
less than the maximum energy attainable with
shock acceleration $e u_{\rm s} B L/c$. This
means that deflection is not a problem for the
highest energy cosmic rays, but would be a
problem for lower energy CR, and, to compute the
cosmic ray current, we are entitled to figure in
only CRs of the highest energies, i.e.\ above
$u_{\rm rms}E_{\rm max}/u_s$. Magnetic field
amplification that uses CRs at lower energies is
unlikely to help increase the maximum energy to
which CRs can be accelerated by shocks.

As mentioned above, the correlation length $L$ is
probably considerably less than the size of the
box, i.e.\ the radius of the supernova remnant
when applied to that context.
It is possible that the rather large values of
magnetic fields that have been claimed for young
supernova remnants ($10^3\, G$, i.e.\ about
$10^2$ times the interstellar field of the
Galaxy) are observationally compatible with such
a small scale.

\subsection{Weakly Magnetized Relativistic Shock Waves}

We define a weakly magnetized relativistic shock wave as one where the
kinetic energy of the upstream fluid flowing into the shock greatly
exceeds the magnetic energy, i.e.\ where the upstream fluid
is sufficiently magnetized as to be describable by MHD.
Particles reflecting off the shock have velocity $\beta_s$ in the frame
of the shock, where $u_s=\beta_s c$ is the shock velocity
which we assume to be equal to the streaming velocity discussed in
\Sec{Introduction}.
Assuming
$(1-\beta_s)\ll 1$, velocities of up to  $\beta_s +(1-\beta_s)/2$ in the
lab frame.
The thickness of the shock precursor in the lab frame is thus of order
$R_s(1-\beta_s)/2$, where $R_s$ is the radius of the shock.
The current density in reflected ions, which we assume extend further
upstream than the reflected electrons, is thus of order
\begin{equation}
J \simeq 2 e n_0 c \, \Gamma_s^2.
\end{equation}
For typical GRB parameters $n_0\sim 1  \rm cm^{-3}$, $u_s \simeq c$,
$\Gamma_{s2}\equiv \Gamma_s/100$ (where $\Gamma_s$ is the shock Lorentz factor),
$B_0\ \sim 3\mu$G, the dimensionless
parameter defined in the introduction is ${\cal J} \sim n_0 m_i c^2/(B_0^2/8\pi)
= \Gamma_s^2 c^2/{\vA}^2  \sim 10^{13}\Gamma_{s2}^2$ for scales $k^{-1}$
of order the reflected ion gyroradius.
This is a far larger value than anything that can be reliably simulated,
because the fastest growing mode is of too small a scale to be resolvable
numerically.

The thickness of the precursor of reflected ions from the shock is
determined by how far ahead of the shock the ions can get before they
are overtaken by the shock. A reflected ion by definition moves  faster
along the shock normal than the shock itself at the moment the ion
crosses upstream, and it is overtaken by the shock after it has gyrated
approximately  $1/\Gamma_s$ of its gyroradius
$r_g \simeq \Gamma_s^2
m_i c^2/e B_0$, at which point its motion along the shock normal is less
than that of the shock, so the shock overtakes it.
Thus, it has moved a distance of $\Delta r = r_g/\Gamma_s= \Gamma_s
m_ic^2/eB_0 $, whereas the shock has moved by $\beta_s \Delta r$.
The thickness of the precursor is then
$(1-\beta_s)\Delta r \simeq m_i c^2/\Gamma_s eB_0$.

The time over which a parcel of fluid at radius $R$ is exposed to the
reflected ion flux is
\begin{equation}
\Delta t = \Delta r(1-\beta_s)/c,
\end{equation}
and its ratio to the Alfv\'en crossing time across the precursor is
\begin{equation}
{\Delta t\over R/\vA}= \vA/c.
\end{equation}

Due to numerical limitations, it remains unclear how much magnetic field
amplification can take place.
While we have shown that perturbations can grow by many orders of
magnitude, we do not have any runs in which the final magnetic field
was more than a factor of 10 or so more than the original field.
As we do not understand the nonlinear dissipation of magnetic field,
so we do not know  {\it  a priori} how the  limiting field scales with
${\cal J}$.
Yet we argue on theoretical grounds that the $\alpha$-dynamo should be
able to amplify the field at relativistic or near relativistic shocks by
a large factor, as it apparently does in many compact objects, and the
existence of such a dynamo is ultimately due to the left--right asymmetry
of the magnetic turbulence that is generated by the cosmic rays.

\subsection{Cosmic Rays in the Galaxy}

Let us now consider systems with lifetimes that are large compared
to the Alfv\'en crossing time, such as the Galaxy. Here there is
enough time for the standard $\alpha \Omega$ mean-field dynamo
dynamo to work if there is a source of right-left asymmetric
turbulence \citep{RSS88,BS05}. In any system where the magnetic
energy has attained rough equipartition between magnetic field and
cosmic ray pressure, the ratio on the right hand side of
\Eq{KeyParameter} obeys $8\pi P^{\rm cr} / B^2 \lesssim 1$.
The first ratio on the right hand side of \Eq{KeyParameter}
$u_s/c$, must also be less than unity.
The third ratio in \Eq{KeyParameter} must also be less than unity
for the assumption for the Bell instability (small deflection of the
current-bearing cosmic rays) to be valid.
It follows that for the above assumptions, the left hand side ${\cal
J} = 4\pi J^{\rm cr} / c k_1 B_0 \ll 1$. This, however, means that
the Bell instability does not occur, as seen from \Eq{A7}; rather,
the effect of the cosmic rays is to  slightly decrease the phase
velocity of stable Alfv\'en waves. It follows that Bell turbulence
cannot amplify the magnetic field to near equipartition with the
cosmic ray pressure. Some estimates in the literature use a cosmic
ray density of $\sim 10^9$ cm$^{-3}$. It would follow that ${\cal J}
\gg 1$. However, the problem here is that this value for the cosmic
ray density includes low energy (GeV) cosmic rays, which satisfy the
assumption for small deflection only at extremely small spatial
scales. It is doubtful  that such modes should even be Bell-unstable
at all because of the $-\eta k^2$ damping term in the imaginary part
of the frequency, as expressed by \Eq{A7}, which grows with $k$
faster than the growth term, and which should therefore dominate at
small spatial scales.

As discussed in the introduction,
the factor $3P^{\rm cr} / (B^2/4\pi)$ is of the order of
unity for the Galaxy as a whole if all relativistic CRs are included.
The anisotropy  $u_s/c$ is bounded by observations to be
at most $10^{-3}$. Finally, the term $eB/ k\Gamma m_i c^2$
must be less than unity to satisfy the small CR deflection
criterion that is the basis for the Bell instability.
Altogether it follows that the quantity ${\cal J}$
in \Eq{KeyParameter},
is less than unity in the Galactic disk.
This, however, implies Bell stability by \Eq{A7}.
So, whereas
the standard $\alpha \Omega$ mean-field dynamo \citep{RSS88,BS05} and
tangling the Galactic magnetic field with
cosmic ray streaming instabilities may be viable ways
to amplify the Galactic magnetic field,
resonant CR streaming instabilities seem the more promising way to do it over
large volumes, where $3P^{\rm cr} / (B^2/4\pi)$
is of the order of unity.
Resonant CR streaming instabilities produce Alfv\'en modes
of preferential circular polarization, just as the Bell instability
does, so the theoretical mechanisms for field amplification that are
discussed in this paper apply to them as well.

We conclude that dynamo activity in the Galactic disk from the twisting
of the field by collective cosmic ray interactions can take place in
principle.
The numbers seem  marginal, so more careful investigation is needed to
settle this point.

\subsection{Unconfined intergalactic cosmic rays}

In intergalactic space, the cosmic ray pressure
is comparable to the magnetic pressure and there
is then the chance that ${\cal J}$ greatly
exceeds unity. Assuming that streaming
instabilities keep the streaming velocity below
the Alfv\'en  velocity $\vA$, we can replace
$u_s$ with $\vA$, so in principle the field can
be amplified to the point where the magnetic
pressure $B^2/4\pi$ is of the order of $P^{\rm
cr}\vA/c$.

We use the following parameters for plasma and
cosmic ray particles: we assume $u^{\rm cr} = 3
\times (10^{6}$--$10^{7})\cm\s^{-1}$ for the
drift velocity of cosmic ray particles, $n^{\rm
cr} = 10^{-9}$--$10^{-10}\cm^{-3}$ for the number
density of cosmic rays particles, $n_i
= 10^{-4}$--$10^{-2}\cm^{-3}$ for the mean number
density of plasma ions, $B_\ast = 1
\,\mu$G for the equilibrium mean magnetic field,
and $\ell_0 = 3 \times 10^{18}$ cm for the
turbulent length scale. The Alfv\'en speed is
$\vA = 10^{8}$--$10^{9}\cm\s^{-1}$ and the
dimensionless ${\cal J}$ parameter is
\begin{equation}
{4\pi\over c} {J^{\rm cr} \ell_0 \over B_\ast}
\approx 10^{2}\mbox{--}10^{5},
\end{equation}
which is comparable to the values studied in this paper.

\section{Conclusions}

We have investigated the mean-field dynamo
mechanisms in a turbulent plasma with a cosmic
ray current. We find a linear growth stage,
corresponding to the Bell instability, and then
a nonlinear stage corresponding to the
production of a fully-developed
MHD turbulence and generation of larger-scale
magnetic field.

In the nonlinear stage, the level of MHD turbulence continues to grow
(see Figs.~\ref{pspec_compt2_tot}
and \ref{ppower_all_EK_EM_T512eo4x_powerbz}), and
the correlation length increases with time (see
\Fig{ppower_kf_T512eo4x_powerbz}).
The turbulence develops a $E_{\it M}(k)\propto k^{-5/3}$ spectrum,
despite the fact that the initial
spectrum is $E_{\it M}(k)\propto k^{4}$.
These effects were far from obvious, even
given the stirring by the counter-$J^{\rm cr}
\times B$ forces on the fluid, which is the basis
of the Bell instability.

We suggest that this combination of magnetic
field amplification and large-scale ordering is
due to the $\alpha^2$ dynamo instability, in
which the original field is stretched by unstable
Alfv\'en modes, and then the stretched component is
itself stretched by unstable circularly polarized
Alfv\'en modes.
This effect is to be contrasted with the long wavelength linear instability
discussed by \cite{BOE10}, because their growth rate vanishes when the
perturbations are independent of $z$.
According to the simulations, the level
of magnetic field on any scale can be enhanced by a factor of several
within one expansion time, even at the largest scales, but there is no
direct numerical evidence at present that it can be enhanced by much
more than that.
Nevertheless, the analysis presented here predicts that much larger
enhancement, via exponential growth to the $\alpha$ effect, is possible.

We have also used DNS and the test-field method to confirm our analysis.
DNS shows that the instability has three stages.
In the first stage, the Bell instability is excited;
in the intermediate stage the linear growth continues among the
high $k$ modes, while mode coupling feeds the low $k$, large-scale modes.
In the third stage, growth on large scale continues,
apparently due to the $\alpha^2$ dynamo.

We find, as expected, that the value of $\alpha$,
which has units of velocity, can never be much
greater than the rms turbulent velocity, $u_{\rm rms}$.
The rms turbulent velocity, $u_{\rm rms}$,
by energy conservation, must be
less than the shock velocity $u_s$, probably much less. Thus the
maximum scale to which the $\alpha^2$ dynamo can
operate efficiently must be limited to
$\alpha/u_s \ll 1$ of the radius of the blast wave $R$;
because $\alpha \lesssim u_{\rm rms}\ll u_s$,
there is not enough time in
a single expansion time to stretch or move the
field significantly over the characteristic
expansion time $R/ u_s$ of the blast wave. There
is enough time only to greatly amplify the field
on scales much smaller than $R$. Similar remarks
would apply to accretion shocks, since material
is swept out of the region of field amplification
over a timescale of $R/u_s$.

In blast waves from GRBs, on the other hand,
this maximum scale of amplification, $\alpha \Delta /c$
(where $\Delta$ is the scale of the cosmic ray precursor),
could be much larger than the ion
skin depth, so dynamo activity in the precursors of such blast waves
could significantly increase the magnetic correlation length
relative to what is produced by the Weibel instability, so that the Ohmic
dissipation downstream would not be so devastating.

The maximum energy to which cosmic rays can be
accelerated by expanding supernova remnants,
which is proportional to $B_0 R$ without magnetic
field amplification, is not greatly enhanced by
large-scale field amplification (not, at least,
in the parameter regime we have explored), since
the increase in field strength comes at the cost
of decreased correlation length $L$, and the
product $B L$ is enhanced far less than $B$.
However, quantification of this point, made in
greater detail by \cite{Eichler-and-Pohl-2011}, would
require a more extensive library of simulations,
for all values of the ion current parameter
${\cal J}$ that could conceivably occur in
nature, and such simulations become very
difficult  for high ${\cal J}$. Clearly if the
pressure of initial field is sufficiently small
relative to the cosmic ray pressure, then
extremely large values of ${\cal J}$ are
possible, and the range ${\cal J} \gg 10^3$ has
not been explored numerically. On the basis of
extrapolation of the two values of ${\cal J}$
(${\cal J} = 80$ or $800$), for which we run
high-resolution simulations, the $e$-folding
timescale for field amplification $t({\cal J},
k)$ over scale $\pi k^{-1}$ is approximately
$C[\vA k {\cal J}^{\delta}]^{-1}$; ($C\sim 1,
\delta \sim 0.7$), while we expect on the basis
of \Eq{KeyParameter} that the largest
allowable value of ${\cal J}$, which occurs at
$eB/ [k\Gamma m_i c^2]\sim 1$,  scales as
$\vA^{-2}$. So the timescale for field
amplification scales as $\vA^{-1+2\delta}k^{-1}
\sim \vA^{0.4}k^{-1}\propto B_0^{0.4}k^{-1}$.
Clearly this timescale can be made small enough
for sufficiently weak fields and sufficiently
small spatial scales (i.e.\ high $k$). How field
amplification at these scales and field strengths
would ultimately affect the capability of shock
acceleration remains an important question for future research.

We have also considered the potential of this
dynamo process in  steady situations where newly
created magnetic flux does not quickly sweep
through the region where cosmic rays stream. An
example of this could be the Galaxy itself, with
escaping CR providing a steady flux over the
lifetime of the Galaxy. If we assume that the CRs
that provide current for growth on scale $k^{-1}$
must satisfy $eB/[k\Gamma m_i c^2]\le 1$, then, by
\Eq{KeyParameter}, the condition that
${\cal J} \ge 1$ requires that $P^{\rm cr}\gg
(B^2/ 8\pi)(c/u_s)$ (where $u_s$ now stands for
CR streaming velocity), so the  dynamo can only
bring the magnetic pressure to  within a fraction
$u_s/c$ of the CR pressure.  On the other hand,
this may be interesting, because it could
``sprout" a seed magnetic field to a sufficiently
large amplitude that some other mechanism, such
as the magnetorotational instability, could
further raise the field pressure to its present
value, $\sim P^{\rm cr}$.

We have also noted that resonant CR streaming
instability, coupled with differential rotation,
remains a possible way to promote an $\alpha \Omega$
dynamo in the Galactic disk.

\begin{acknowledgements}
We thank Yuri Lyubarsky for helpful discussions.
We acknowledge the NORDITA dynamo program of 2011
for providing a stimulating scientific
atmosphere. This work was supported in part by
the Swedish Research Council Grant No.\
621-2007-4064, the European Research Council
under the AstroDyn Research Project 227952 (AB)
and by the Israel Science Foundation governed by
the Israeli Academy of Sciences, the Israel-U.S.
Binational Science Foundation, and the Joan and
Robert Arnow Chair of Theoretical Astrophysics.
(DE). The authors (IR and NK) acknowledge the
hospitality of NORDITA.
\end{acknowledgements}

\appendix

\section{Derivation of equation for total $\alpha$ effect}

We determine the contributions to the mean electromotive
force, $\bec{\cal E}(\overline{\bm B}) = \overline{ {\bm u}
{\bm \times} {\bm b} }$, caused by cosmic ray
particles in homogeneous turbulent plasma.
The procedure of the derivation of equation for the mean
electromotive force is as follows.
We use a mean field approach in which the magnetic and
velocity fields are divided into the mean and fluctuating
parts, where the fluctuating parts have zero mean values.
The momentum and induction equations for the turbulent
fields are given by
\begin{eqnarray}
\overline{\rho} \, {\partial {\bm u}(t, {\bm x}) \over
\partial t} &=& - \bec{\nabla} p_{\rm tot} + {1 \over
4\pi} [({\bm b} {\bm \cdot}
\bec{\nabla}) \overline{\bm B} + (\overline{\bm B}
{\bm \cdot} \bec{\nabla}){\bm
b}]
- {1\over c}\overline{{\bm J}^{\rm cr}} {\bm \times} {\bm b}
+ {1\over c} e n^{\rm cr} \, \left({\bm u} {\bm \times}
\overline{\bm B}\right) + {\bm N}^u  ,
\label{B1} \\
{\partial {\bm b}(t,{\bm x}) \over \partial t} &=&
(\overline{\bm B}
{\bm \cdot} \bec{\nabla}){\bm u} - ({\bm u} {\bm \cdot}
\bec{\nabla}) \overline{\bm B} - \overline{\bm B}
\, (\bec{\nabla} {\bm \cdot} {\bm u}) + {\bm N}^b ,
\label{B2}
\end{eqnarray}
where ${\bm u}$ and ${\bm b}$ are fluctuations of
velocity and
magnetic field, $\overline{\bm B}$ is the mean magnetic
field, $\overline{{\bm J}^{\rm cr}}$ is the mean density
of the electric current of cosmic ray particles,
${\bm N}^{u}$ and ${\bm N}^{b}$ are the nonlinear
terms which include the molecular dissipative terms,
$ p_{\rm tot} = p + (\overline{\bm B}
{\bm \cdot} {\bm b})/4\pi$ are the fluctuations of total
pressure, $p$ are the fluctuations of fluid pressure.
To exclude the pressure term from the equation of
motion~(\ref{B1}) we calculate $\bec{\nabla} {\bm
\times} (\bec{\nabla} {\bm \times} {\bm u})$.
Then we rewrite the obtained equation and \Eq{B2}
in Fourier space.

\subsection{Two-scale approach}

We apply the two-scale approach,
e.g., a correlation function,
\begin{eqnarray*}
\overline{ u_i ({\bm x})
u_j ({\bm  y}) } &=& \int \,d{\bm k}_1 \, d{\bm  k}_2 \,
\overline{ u_i({\bm  k}_1) u_j ({\bm k}_2) }
\exp \{i({\bm  k}_1 {\bm
\cdot} {\bm x} + {\bm  k}_2 {\bm \cdot} {\bm y})
\}  = \int \,d {\bm  k} \,d {\bm  K} \, f_{ij}({\bm k, K})
\exp (i {\bm k} {\bm \cdot} {\bm r}+ i {\bm K}
{\bm \cdot} {\bm R})
\nonumber \\
&=& \int  \,d {\bm  k} \, f_{ij}({\bm k, R}) \exp
(i {\bm k} {\bm \cdot}{\bm r}) ,
\end{eqnarray*}
\citep[see, e.g.,][]{RS75}.
Hereafter we omitted argument $t$ in the correlation
functions, $f_{ij}({\bm k, R}) = \hat L(u_i; u_j)$,
where
\begin{eqnarray*}
\hat L(a; c) = \int \overline{ a({\bm k}
+ {\bm  K} / 2) c(-{\bm k} + {\bm  K} / 2)
} \exp{(i {\bm K} {\bm \cdot} {\bm R}) }
\,d {\bm  K},
\end{eqnarray*}
and we introduced new variables
${\bm R} = ({\bm x} +  {\bm y}) / 2,$
$\, {\bm r} = {\bm x} - {\bm y}$, $\, {\bm K}
= {\bm k}_1 + {\bm k}_2$, ${\bm k} = ({\bm k}_1
- {\bm k}_2) / 2$.
The variables ${\bm R}$ and ${\bm K}$ correspond
to the large scales, while ${\bm r}$ and ${\bm k}$
correspond to the  small scales.
This implies
that we assumed that there exists a separation of
scales, i.e.,
the maximum scale of turbulent motions $\ell_0$ is
much smaller
than the characteristic scale $L_B$ of inhomogeneity
of the mean
magnetic field.

\subsection{Equations for the second moments}

We derive equations for the following correlation
functions:
$f_{ij}({\bm k, R}) = \hat L(u_i; u_j)$,
$\, h_{ij}({\bm k, R})
= (4\pi \, \overline{\rho})^{-1} \, \hat L(b_i; b_j)$
and $g_{ij}({\bm k, R}) = \hat L(b_i; u_j)$.
The equations for these correlation functions are given by
\begin{eqnarray}
&& {\partial f_{ij}({\bm k}) \over \partial t}
= i({\bm k} {\bm
\cdot} \overline{\bm B}) \Phi_{ij}
+ D_{im}({\bm k}_1)f_{mj}
+ D_{jm}({\bm k}_2)f_{im}
+ A_{im}({\bm k}_1) g_{mj}({\bm k})
+ A_{jm}({\bm k}_2)g_{im}(-{\bm k}) + I^f_{ij}
+ f_{ij}^N  ,
\label{B6} \\
&& {\partial h_{ij}({\bm k}) \over \partial t}
= - i({\bm k}{\bm
\cdot} \overline{\bm B}) \Phi_{ij}
+ i k_n[g_{in}({\bm k}) \, \overline{B}_j
- g_{jn}({\bm k})\, \overline{B}_i]
+ I^h_{ij} + h_{ij}^N ,
\label{B7} \\
&& {\partial g_{ij}({\bm k}) \over \partial t}
= i({\bm k} {\bm
\cdot} \overline{\bm B}) [f_{ij}({\bm k})
- h_{ij}({\bm k}) -
h_{ij}^{(H)}] - i k_m \overline{B}_i \, f_{mj}
+ D_{jm}({\bm k}_2)g_{im}({\bm k})
+(4 \pi \overline{\rho})\,A_{jm}({\bm k}_2)h_{im} + I^g_{ij}+ g_{ij}^N ,
\label{B8}
\end{eqnarray}
where $\Phi_{ij}({\bm k}) = (4\pi \,
\overline{\rho})^{-1} \, [g_{ij}({\bm k})
- g_{ji}(-{\bm k})]$.
Hereafter we omitted argument ${\bm R}$ in the
correlation functions and neglected terms
$\sim O(\nabla_{\bm R}^2)$,
\begin{eqnarray}
D_{ij} = 2 \varepsilon_{ijp} \tilde\Omega_q^{\rm cr}
k_{pq} \,, \quad A_{ij} = 2 \varepsilon_{ijp}
J_q^{\rm cr} k_{pq} \,, \quad
\tilde{\bm \Omega}^{\rm cr} = {e n^{\rm cr}
\overline{\bm B} \over 2 c \, \overline{\rho}} \,,
\quad {\bm J}^{\rm cr} = {\overline{{\bm J}^{\rm cr}}
\over 2 c \, \overline{\rho}} ,
\label{B10}
\end{eqnarray}
$\varepsilon_{ijn}$ is the fully antisymmetric
Levi-Civita tensor, the terms $f_{ij}^{N} ,$
$\, h_{ij}^{N} $ and $ g_{ij}^{N} $ are determined
by the third moments appearing due to the nonlinear
terms, the source terms $I_{ij}^f$ , $\, I_{ij}^h$
and $I_{ij}^g$ which contain the large-scale spatial
derivatives of the mean magnetic and velocity fields,
are given by Eqs.~(A3)--(A6) in \cite{RK04}.
These terms determine turbulent magnetic diffusion
and effects of nonuniform mean velocity on mean
electromotive force.
In the present study we neglect small effects of
cosmic ray particles on the turbulent magnetic
diffusion.

For the derivation of Eqs.~(\ref{B6})--(\ref{B8})
we use an approach that is similar to that
applied in \cite{RK04}.
We took into account that the terms with symmetric
tensors with respect to the indexes $i$ and $j$
in \Eq{B8} do not contribute to the mean
electromotive force because ${\cal E}_{m} =
\varepsilon_{mji} \, g_{ij}$.
We split all tensors into nonhelical, $h_{ij},$
and helical, $h_{ij}^{(H)},$ parts.
The helical part of the tensor of magnetic
fluctuations $h_{ij}^{(H)}$ depends on the
magnetic
helicity, and the equation for $h_{ij}^{(H)}$
follows from the magnetic helicity conservation
arguments \citep[see, e.g.,][and references therein]{RK04,BS05}.

\subsection{$\tau$-approach}
\label{Tau}

The second-moment equations~(\ref{B6})--(\ref{B8})
include the first-order spatial differential
operators $\hat{\cal N}$  applied to the third-order
moments $M^{\rm(III)}$.
A problem arises how to close the system, i.e.,
how to express the set of the third-order terms
$\hat{\cal N} M^{\rm(III)}$ through the lower
moments $M^{\rm(II)}$.
We use the spectral $\tau$ approximation which
postulates that the deviations of the third-moment
terms, $\hat{\cal N} M^{\rm(III)}({\bm k})$, from
the contributions to these terms afforded by
the background turbulence, $\hat{\cal N}
M^{\rm(III,0)}({\bm k})$, are expressed through
the similar deviations of the second moments:
\begin{eqnarray}
&& \hat{\cal N} M^{\rm(III)}({\bm k}) - \hat{\cal N}
M^{\rm(III,0)}({\bm k})
=- {1 \over \tau(k)} \,  \left[M^{\rm(II)}({\bm k})
- M^{\rm(II,0)}({\bm k})\right] ,
\label{T10}
\end{eqnarray}
\citep{O70,PFL76,KRR90,RK04}, where $\tau(k)$ is
the scale-dependent relaxation time, which
can be identified with the correlation time of the
turbulent velocity field.
The quantities with the superscript $(0)$ correspond
to the background turbulence (see below).
We apply the spectral $\tau$ approximation only
for the nonhelical part $h_{ij}$ of the tensor of
magnetic fluctuations.

\subsection{Solution of equations for the second moments}

First we solve Eqs.~(\ref{B6})--(\ref{B8}) neglecting
the sources $I^f_{ij}, I^h_{ij}, I^g_{ij}$ with the
large-scale spatial derivatives.
Then we will take into account the terms with the
large-scale spatial
derivatives by perturbations.
We subtract from Eqs.~(\ref{B6})--(\ref{B8}) the
corresponding equations written for the background
turbulence, use the spectral $\tau$ approximation
and neglect the terms with the large-scale spatial
derivatives.
We assume that the cosmic ray velocity is much
larger than fluid velocity, so that the terms
$\propto D_{ij}$ in Eqs.~(\ref{B6})--(\ref{B8}) vanish.
Next, we neglect the effect related to the compressibility
of the turbulent velocity field. Such effects are important
when the Mach number is of the order of or larger than 1.
We also assume that the characteristic time of variation
of the second moments is substantially larger than the
correlation time $\tau(k)$ for all turbulence scales.
This allows us to get a stationary solution for the
equations for the second-order moments, $M^{\rm(II)}$.
Thus, we arrive to the following steady-state solution of
Eqs.~(\ref{B6})--(\ref{B8}):
\begin{eqnarray}
&& \hat f_{ij}({\bm k}) \approx f_{ij}^{(0)}({\bm k})
+ i \tau
({\bm k} {\bm \cdot} \overline{\bm B}) \hat
\Phi_{ij}({\bm k})
+ \tau \, \Big[A_{im}\, \hat g_{mj}({\bm k}) + A_{jm} \,
\hat g_{im}(-{\bm k})\Big] ,
\label{B17}\\
&& \hat h_{ij}({\bm k}) \approx h_{ij}^{(0)}({\bm k})
- i \tau({\bm k} {\bm \cdot} \overline{\bm B}) \hat
\Phi_{ij}({\bm k}) ,
\label{B18}\\
&& \hat g_{ij}({\bm k}) \approx g_{ij}^{(0)}({\bm k})
+ i \tau ({\bm k} {\bm \cdot} \overline{\bm B})
\left[\hat f_{ij}({\bm k}) - \hat h_{ij}({\bm k})\right]
+ \tau \, (4 \pi \overline{\rho}) \, A_{jm} \, \hat h_{im}({\bm k}),
\label{B19}
\end{eqnarray}
where $\hat f_{ij}, \hat h_{ij}$ and $\hat g_{ij}$ are
solutions
without the sources $I^f_{ij}, I^h_{ij}$ and $I^g_{ij}$.
In the present study we consider linear effects in
perturbations of the mean magnetic field.
The nonlinear mean-field modeling in turbulent compressible MHD
flows with cosmic rays is a subject of a separate ongoing study.

\subsection{Model for the background turbulence}

Now we need a model for the background anisotropic turbulence
[see Eqs.~(\ref{B17})--(\ref{B19})].
The anisotropy is caused by the equilibrium mean
magnetic field $\overline{\bm B}_\ast$.
Generally, a model of an anisotropic turbulence with
one preferential direction can be constructed as a
combination of three-dimensional isotropic turbulence and
two-dimensional turbulence in the plane perpendicular
to the preferential direction \citep[see, e.g.,][]{EKRZ02}.
Also we take into account that the tensor
$f_{ij}^{(0)}({\bm k})$ is the sum of non-helical
and helical parts of the turbulence.
A non-zero kinetic helicity is caused by the Bell instability.
To relate the velocity fluctuations tensor
$f_{ij}^{(0)}({\bm k})$ with the magnetic fluctuations
tensor $h_{ij}^{(0)}({\bm k})$ and the cross-helicity tensor
$g_{ij}^{(0)}({\bm k})$, we use the relation between
the magnetic and the velocity fields in the Bell mode:
${\bm b}^{(0)}({\bm k})= i ({\bm k} {\bm \cdot}
{\bm B}_\ast) \,{\bm u}^{(0)}({\bm k}) / \gamma_{\it B}$,
where $\gamma_{\it B}$ is determined by \Eq{A7}.
We use the following model for the background anisotropic
homogeneous and helical turbulence caused by the
Bell instability:
\begin{eqnarray}
f_{ij}^{(0)}({\bm k}) &\equiv& \overline{ u^{(0)}_i({\bm k})
\, u^{(0)}_j(-{\bm k}) } = {E(k) \over 8 \pi k^2}
\left\{\left[(1 - \epsilon) \, (\delta_{ij} - k_{ij})
+ 2 \epsilon \, \left(\delta_{ij}  - e_{i} e_{j} - k^{\perp}_{ij} \right)\right]
\, \overline{[{\bm u}^{(0)}]^2}
- {i \over k^2} \, \varepsilon_{ijn} \, k_n
\, \overline{{\bm u}^{(0)} \cdot (\bec{\nabla} \times {\bm u}^{(0)})} \right\} ,
\nonumber\\
\label{B11}\\
h_{ij}^{(0)}({\bm k}) &\equiv& {\overline{ b^{(0)}_i({\bm k})
\, b^{(0)}_j(-{\bm k}) } \over 4\pi
\overline{\rho}}= \left(L^{\rm cr} \, k\right)
\, f_{ij}^{(0)}({\bm k}) ,
\label{B12}\\
g_{ij}^{(0)}({\bm k}) &\equiv& \overline{ b^{(0)}_i({\bm k})
\, u^{(0)}_j(-{\bm k})
} = \left(4 \pi \overline{\rho} \,
L^{\rm cr} \, k\right)^{1/2} \, {i ({\bm k} {\bm \cdot}
\overline{\bm B}_\ast) \over |{\bm k} {\bm \cdot}
\overline{\bm B}_\ast|} \, f_{ij}^{(0)}({\bm k}) ,
\label{B13}
\end{eqnarray}
where $L^{\rm cr} = c \, \overline{B}_\ast /
(4 \pi \overline{J^{\rm cr}})$, $\, E(k) = (q-1) \,
\ell_{0} \, (\ell_{0} \, k)^{-q}$ is the energy
spectrum function, the length $\ell_{0}$ is the
maximum scale of turbulent motions, ${\bm e}$
is the unit vector directed along the equilibrium
mean magnetic field $\overline{\bm B}_\ast$,
$\, \delta_{ij}$ is the Kronecker unit tensor,
$0 < \epsilon \leq 1$ is the anisotropy parameter
of turbulence, ${\bm k} = {\bm k}^{\perp} + k_z
\, {\bm e}$, $\, k_z = ({\bm k} {\bm\cdot} {\bm e})$,
$k_{ij}=k_{i} k_{j} / k^2$ and $k^{\perp}_{ij}
= k^{\perp}_{i} k^{\perp}_{j} / ({\bm k}^{\perp})^{2}$.
The turbulent correlation time is $\tau(k)
= C_\tau \, \tau_0 \, (\ell_{0} \, k)^{-\mu}$,
where the time
$\tau_0 = \ell_{0} / u_0$, $\, u_0=\sqrt{\overline{
[{\bm u}^{(0)}]^2 }}$ is the characteristic turbulent
velocity in the scale $\ell_{0}$ and
$C_\tau$ is the coefficient.
For the background turbulence with a constant
dissipation rate of turbulent energy in inertial
range of scales, the exponent $\mu=q-1$,
the energy spectrum $E(k) \propto - d\tau / dk$
and the coefficient $C_\tau=2$.

Using the solution of the derived second-moment equations
(\ref{B17})--(\ref{B19}), we determine
the contributions to the mean electromotive force,
${\cal E}_{i}^{\rm cr} = \varepsilon_{imn} \,
\int \overline{ b_n({\bm k}) \, u_m(-{\bm k}) }
\,{\rm d} {\bm k}$, caused by cosmic ray
particles in homogeneous turbulent plasma.

\subsection{Derivations of contributions to the $\alpha$ effect}
\label{Alpha}

We take into account effects which are linear in
the perturbations of the mean magnetic field:
$\tilde{\bm B}=\overline{\bm B} -\overline{\bm B}_\ast$,
i.e., we consider a kinematic mean-field dynamo.
Substituting Eqs.~(\ref{B17})--(\ref{B18})
into \Eq{B19} we obtain:
\begin{eqnarray}
\hat g_{ij}({\bm k}) &\approx& \hat g_{ij}^{(I)}({\bm k})
 +\hat g_{ij}^{(II)}({\bm k}) + \hat g_{ij}^{(III)}({\bm k}),
\label{B60}\\
\hat g_{ij}^{(I)}({\bm k}) &\approx& i \tau ({\bm k} {\bm \cdot} \tilde{\bm B})
\left[\hat f_{ij}^{(0)}({\bm k}) - \hat h_{ij}^{(0)}({\bm k})\right],
\label{B61}\\
\hat g_{ij}^{(II)}({\bm k}) &\approx& i \tau^2 ({\bm k} {\bm \cdot}
\tilde{\bm B})\Big[A_{im}\, \hat g_{mj}^{(0)}({\bm k}) + A_{jm} \,
\hat g_{im}^{(0)}(-{\bm k}) - (4 \pi \overline{\rho}) \, A_{jm}
\, \Phi_{ij}^{(0)}({\bm k})\Big]
\nonumber\\
&=& i \tau^2 ({\bm k} {\bm \cdot} \tilde{\bm B})\Big[A_{im}\,
\hat g_{mj}^{(0)}({\bm k}) -3 A_{jm} \,
\hat g_{mi}^{(0)}({\bm k})\Big],
\label{B62}\\
\hat g_{ij}^{(III)}({\bm k}) &\approx& g_{ij}^{(0)}({\bm k})
+ \tau \, (4 \pi \overline{\rho}) \, A_{jm} \, h_{ij}^{(0)}({\bm k}),
\label{B63}
\end{eqnarray}
where we have taken into account that $\hat
g_{ij}^{(0)}({\bm k})= \hat g_{ji}^{(0)}({\bm
k})=-\hat g_{ij}^{(0)}(-{\bm k})$. The mean
electromotive force is given by ${\cal
E}_{i}^{\rm cr} = \varepsilon_{imn} \, \int \hat
g_{nm}({\bm k})\,{\rm d} {\bm k}$, where the
tensor $\hat g_{ij}({\bm k})$ is determined by
\Eq{B60}. There are two contributions to
the $\alpha$ effect caused by:
\begin{itemize}
\item[(i)] non-zero kinetic helicity produced by
Bell-instability; this contribution is
determined by the tensor $\hat g_{ij}^{(I)}({\bm k})$, where
the background turbulence $\hat f_{ij}^{(0)}({\bm k})$
is described by the term, $\propto - (i / k^2)
\, \varepsilon_{ijn} \, k_n \, \overline{{\bm u}^{(0)}
\cdot (\bec{\nabla} \times {\bm u}^{(0)})}$
in \Eq{B11}; and
\item[(ii)] interaction of the mean cosmic
ray current with small-scale anisotropic turbulence;
this contribution is determined by the tensor
$\hat g_{ij}^{(II)}({\bm k})$, where
the background turbulence $\hat f_{ij}^{(0)}({\bm k})$
is determined by the term, $\propto \epsilon \,
(\delta_{ij}  - e_{i} e_{j} - k^{\perp}_{ij})$
in \Eq{B11}.
\end{itemize}

The first contribution to the mean electromotive
force caused by a non-zero kinetic helicity
effect in the Bell turbulence is given by
\begin{eqnarray}
{\cal E}_{i}^{(I)} = \varepsilon_{imn} \int \hat g_{nm}^{(I)}({\bm k})
\,{\rm d} {\bm k} = i \varepsilon_{imn}
\, \int \tau(k) \,  ({\bm k} {\bm \cdot}
\tilde{\bm B}) \, \hat f_{nm}^{(0)}({\bm k})
\,{\rm d} {\bm k} = \alpha_{ij}^{(I)} \,
\tilde{B}_j,
\label{B28}
\end{eqnarray}
where $\alpha_{ij}^{(I)}=\alpha^{\rm cr}_1 \, \delta_{ij}$ and
$\alpha^{\rm cr}_1$ is determined by \Eq{D7}.
In the derivation of \Eq{B28} we have taken into account that
$L^{\rm cr} \, k_1 = {\cal J}^{-1} \ll 1$ that allows us to drop
contributions $\propto \hat h_{ij}^{(0)}({\bm k})$ in
comparison with that proportional to $\hat f_{ij}^{(0)}({\bm k})$.

The second contribution to the mean electromotive
force, ${\cal E}_{i}^{(II)}$, is caused by the non-helical part
of the anisotropic turbulence:
\begin{eqnarray}
{\cal E}_{i}^{(II)} = \varepsilon_{imn} \int \hat g_{nm}^{(II)}({\bm k})
\,{\rm d} {\bm k} = 4 \, i \, \varepsilon_{inm}
\, \int \tau^2(k) \, ({\bm k} {\bm \cdot}
\tilde{\bm B}) \, A_{mp} \, g_{np}^{(0)}({\bm k})
\,{\rm d} {\bm k} = \alpha_{ij}^{(II)} \,
\tilde{B}_j,
\label{B29}
\end{eqnarray}
where the tensor $\alpha_{ij}^{(II)}= \alpha^{\rm cr}_2
\, (\delta_{ij} + e_{i} e_{j})$ and $\alpha^{\rm cr}_2$
is determined by \Eq{D4}.
The third contribution to the mean electromotive force,
${\cal E}_{i}^{(III)}=\varepsilon_{imn} \int \hat g_{nm}^{(III)}({\bm k})
\,{\rm d} {\bm k}$, is constant and,
therefore, does not affect the large-scale dynamo.

\subsection{Integrals used in \Sec{Alpha}}

To integrate over the angles in ${\bm k}$
space we used the
following identities:
\begin{eqnarray}
&&\int k_{ijn} \, {\rm sgn}(k_z) \, \sin \theta
\,d\theta \,d\varphi = {\pi \over 2} \biggl[P_{in}({\bm e})
\, e_j + P_{jn}({\bm e}) \, e_i
+ P_{ij}({\bm e}) \, e_n + 2 e_i e_j e_n
\biggr] ,
\label{B21}\\
&&\int {k_{i}^\perp k_{j} k_{n} \over k^3}
\, {\rm sgn}(k_z) \, \sin \theta \,d\theta
\,d\varphi = {\pi \over 2} \biggl[P_{ij}({\bm e})
\, e_n
+ P_{in}({\bm e}) \, e_j\biggr] ,
\label{B23}
\end{eqnarray}
where $P_{in}({\bm e})=\delta_{ij} - e_i e_j$
and $k_z=k \, \cos \theta$.

\subsection{The realizability condition}

Let us consider the case when the spectral functions for
the kinetic helicity, $\chi(k)$, and turbulent kinetic
energy, $u_0^2 \, E(k)$, are different, where
$\overline{{\bm u}^{(0)} \cdot (\bec{\nabla} \times
{\bm u}^{(0)})} = \int \chi(k) \, dk$ and
$\overline{[{\bm u}^{(0)}]^2}=u_0^2 \int E(k) \, dk$.
The realizability condition for the kinetic helicity
\citep{M78} reads:
\begin{eqnarray}
\chi(k) \leq u_0^2 \, E(k) k .
\label{B70}
\end{eqnarray}
Let us determine the explicit expression for the function
$\chi(k)$ using \Eq{EX11} for the estimate for the
the kinetic helicity $\overline{{\bm u}^{(0)} {\bf \cdot}
\left(\bec{\nabla} \times {\bm u}^{(0)}\right)}$ for the
Bell background turbulence:
\begin{eqnarray}
\overline{{\bm u}^{(0)} {\bf \cdot} \left(\bec{\nabla}
\times {\bm u}^{(0)}\right)} \propto
{\tau \, \overline{J_j^{\rm cr}} \over c \,
\overline{\rho}} \, \overline{u_n^{(0)} \nabla_j b_n^{(0)}} =
\left({4\pi\over c}{\overline{J^{\rm cr}} \, \ell_0 \over
\overline{B}_\ast} \right)^{1/2}
\, \overline{V}_A \, {u_0^2 \over 2\ell_0^2}
\int \left(\ell_0 \, k \right)^{3/2} \tau(k) \, E(k) \, dk,
\label{B71}
\end{eqnarray}
where $E(k) = (q-1) \,\ell_{0} \, (\ell_{0} \, k)^{-q}$
and $\tau(k) = 2 \, \tau_0 \, (\ell_{0} \, k)^{1-q}$.
Therefore, the function $\chi(k)$ is given by
\begin{eqnarray}
\chi(k) = \sqrt{\cal J}\, \, \overline{V}_A \, {u_0 \over \ell_0}
\left(\ell_0 \, k \right)^{5/2 - q} \, E(k),
\label{B72}
\end{eqnarray}
and the realizability condition for the kinetic helicity yields:
\begin{eqnarray}
\sqrt{\cal J}\, \,  \, {\overline{V}_A \over u_0} \leq
\left(\ell_0 \, k \right)^{q - 3/2},
\label{B73}
\end{eqnarray}
where
\begin{eqnarray}
{\cal J}={4\pi\over c}{\overline{J^{\rm cr}} \, \ell_0 \over \overline{B}_\ast},
\label{B74}
\end{eqnarray}
and the Kolmogorov spectrum corresponds to $q=5/3$.

%r e f

\end{document}